\documentclass[sn-mathphys-num]{sn-jnl}

\usepackage[square,numbers]{natbib}

\usepackage{graphicx}%
\usepackage{multirow}%
\usepackage{amsmath,amssymb,amsfonts}%
\usepackage{amsthm}%
\usepackage{mathrsfs}%
\usepackage[title]{appendix}%
\usepackage{xcolor}%
\usepackage{textcomp}%
\usepackage{manyfoot}%
\usepackage{booktabs}%
\usepackage{algorithm}%
\usepackage{algorithmicx}%
\usepackage{algpseudocode}%
\usepackage{listings}%

\usepackage{bm}        

\def\bs{\boldsymbol}  
\def\s2n{S^{\prime}/N}

\def\gsim{\;\rlap{\lower 2.5pt
\hbox{$\sim$}}\raise 1.5pt\hbox{$>$}\;}
\def\lsim{\;\rlap{\lower 2.5pt
\hbox{$\sim$}}\raise 1.5pt\hbox{$<$}\;}

\newcommand{\araa}{Annu. Rev. Astron. Astrophys.}   
\newcommand{\aj}{Astron. J.}   
\newcommand{\apj}{Astrophys. J.}   
\newcommand{\apjl}{Astrophys. J. Lett.}   
\newcommand{\apjs}{Astrophys. J. Suppl. Ser.}   
\newcommand{\apss}{Astrophys. Space Sci.}   
\newcommand{\aap}{Astron. Astrophys.}   
\newcommand{\aapr}{Astron. Astrophys. Rev.}   
\newcommand{\aaps}{Astron. Astrophys. Suppl.}   
\newcommand{\mnras}{Mon. Not. R. Astron. Soc.}   
\newcommand{\nat}{Nature} 
\newcommand{\pasj}{Publ. Astron. Soc. Jpn}   
\newcommand{\pasp}{Publ. Astron. Soc. Pac.}   
\theoremstyle{thmstyleone}%
\theoremstyle{thmstyletwo}%

\theoremstyle{thmstylethree}%

\begin{document}

\title[Article Title]{
The Formation of Protoplanetary Disks through Pre-Main Sequence Bondi-Hoyle Accretion.
}


\author[1,2]{\fnm{Paolo} \sur{Padoan}}\equalcont{These authors contributed equally to this work.}

\author*[3]{\fnm{Liubin} \sur{Pan}}\email{panlb5@mail.sysu.edu.cn}
\equalcont{These authors contributed equally to this work.}

\author[1,4]{\fnm{Veli-Matti} \sur{Pelkonen}}

\author[5]{\fnm{Troels} \sur{Haugb{\o}lle}}

\author[5]{\fnm{{\AA}ke} \sur{Nordlund}}

\affil[1]{\orgdiv{Institut de Ci\`{e}ncies del Cosmos}, \orgname{Universitat de Barcelona}, \orgaddress{\street{Mart\'{i} i Franqu\'{e}s 1}, \city{Barcelona}, \postcode{E08028}, \country{Spain}}}

\affil[2]{\orgdiv{Department of Physics and Astronomy}, \orgname{Dartmouth College}, \orgaddress{\street{6127 Wilder Laboratory}, \city{Hanover}, \postcode{03755}, \state{NH}, \country{USA}}}

\affil[3]{\orgdiv{School of Physics and Astronomy}, \orgname{Sun Yat-sen University}, \orgaddress{\street{2 Daxue Road}, \city{Zhuhai}, \postcode{519082}, \state{Guangdong}, \country{China}}}

\affil[4]{\orgname{INAF - Istituto di Astrofisica e Planetologia Spaziali}, \orgaddress{\street{Via Fosso del Cavaliere 100}, \city{Roma}, \postcode{I-00133}, \country{Italy}}}

\affil[5]{\orgdiv{Niels Bohr Institute}, \orgname{University of Copenhagen}, \orgaddress{\street{{\O}ster Voldgade 5-7}, \city{Copenhagen}, \postcode{DK-1350}, \country{Denmark}}}


\abstract{Protoplanetary disks are traditionally described as finite mass reservoirs left over by the gravitational collapse of the protostellar core, a view that strongly constrains both disk evolution and planet formation models. We propose a different scenario where protoplanetary disks of pre-main sequence stars are primarily assembled by Bondi-Hoyle accretion from the parent gas cloud. We demonstrate that Bondi-Hoyle accretion can supply not only the mass, but also the angular momentum necessary to explain the observed size of protoplanetary disks. Additionally, we predict how the specific angular momentum of protoplanetary disks scales with stellar mass. Our conclusions are based on a new analytical derivation of the scaling of the angular momentum in turbulent flows, which we confirm with a numerical simulation of supersonic turbulence. A key outcome of our analysis is the recognition that density fluctuations in supersonic turbulence—previously overlooked in studies of cloud and core rotation—lead to a significant increase in angular momentum at disk-forming scales. This revised understanding of disk formation and evolution alleviates several longstanding observational discrepancies and compels substantial revisions to current models of disk and planet formation.}




\maketitle

Theoretical models of protoplanetary disks (PDs) have so far been focused on the myriad of internal disk processes \citep[e.g.][]{Lesur+23,Drazkowska+23}, ignoring the disks' environment and specifically the possibility of mass infall from larger scales. This implicit assumption that PDs are fully formed at the end of the protostellar collapse is unfounded. It compounds observational problems, from the origin of planetary masses \citep{Ansdell+2016,Manara+18,Mulders+21,Stefansson+23} to the disk lifetimes \citep{Fedele+10,Manara+20,Mauco+23}, from the disk angular momentum\citep{Gaudel+20} to the misalignment of disks \citep{Marino+15,Pinilla+18,Ansdell+20} or exoplanetary orbits \citep{Hubert+13,Sanchis-Ojeda+13,Kamiaka+19,Hjorth+21,Albrecht+22,Kawai+23}. It is also incompatible with recent discoveries of streamers connected to young disks \citep{Pineda+20,Alves+20,Huang+21,Grant+21,Valdivia+22,Cacciapuoti+23} and it contradicts theoretical and computational evidence that Bondi-Hoyle (BH) accretion \citep{Hoyle+Lyttleton39,Bondi+Hoyle44,Bondi52} in young pre-main-sequence (PMS) stars may control the mass budget of their disks \citep{Padoan+05_BH,Throop+Bally08,Padoan+14acc}. In support of an alternative scenario where the evolution of PDs is strongly affected by mass infall, first proposed by Padoan et al. (2005) \citep{Padoan+05_BH}, we demonstrate both analytically and numerically that BH accretion is relevant not only to disk masses (as we had already shown \citep{Padoan+05_BH,Padoan+14acc}), but also to their angular momenta or sizes. This scenario leads to predictions of the observed relations between disk properties and stellar mass \citep{Andrews+2018,Hendler+20,Andrews20,Long+22,Testi+22} that remain unexplained in the standard models of isolated disks, as suggested by a recent numerical study of late accretion events \citep{Kuffmeier+23}. 

To date, apart from the BH accretion scenario presented here, no other theoretical model successfully predicts the disk sizes of PMS stars and their dependence on stellar mass. In addition, the scaling of angular momentum in supersonic turbulence, which dictates the angular momentum captured by PMS stars onto their disks through BH accretion, had never been derived prior to this work. As shown here, the angular momentum scaling commonly expected from the velocity-size relation of star-forming clouds--an adequate predictor of their rotation rates \citep[e.g.][]{Goodman+93}--neglects a dominant contribution due to strong density fluctuations. This conventional scaling would severely underestimate the angular momentum captured through BH accretion, leading to the incorrect conclusion that BH accretion could not account for the observed disk sizes. In contrast, our newly derived angular momentum scaling for supersonic turbulence demonstrates that both the disk sizes and their dependence on stellar mass are, in fact, natural outcomes of the BH accretion process.

We consider the specific angular momentum, $\bs j$, of the gas in a sphere of radius $R$ with respect to the center of the sphere, in a turbulent medium with an rms gas velocity $\sigma_{v,0}$ at a large driving scale $R_0$, and evaluate the dependence of its standard deviation, $\langle {\bs j}^2\rangle^{1/2}$, on the scale $R$. There are two distinct contributions to $\bs j$. The first one is due to the offset of the center of mass from the center of the sphere, because of random density fluctuations, so the velocity of the center of mass carries a net angular momentum. The second one is the net rotation of the gas around the center of the sphere because of random velocity fluctuations. In Methods, we demonstrate that these two contributions lead to different scaling laws, depending on which one dominates. 
In incompressible turbulence, the first contribution vanishes because of the constant density, so only the local net rotation from velocity fluctuations across the scale $R$ contributes to the angular momentum, and the scaling of $\langle {\bs j}^2\rangle^{1/2}$ can be derived from the velocity scaling by dimensional analysis. Net rotation is the only contribution also in observational estimates of $j$ in molecular clouds (MCs), where density fluctuations are ignored and the mean cloud velocity is subtracted away, so the contribution from the offset of the center of mass is removed by design (Methods, Section\,\ref{Larson}). As in incompressible turbulence, the scaling of the observed $j$ in MCs follows from dimensional analysis of the velocity scaling. 
Adopting the observational velocity-size relation of MCs (Methods, Section\,\ref{timescale}) yields
\begin{align}
\langle \bs j^2  \rangle^{1/2}  = 8.3 \times 10^{22} (R/1\,{\rm pc})^{1.5} {\rm cm}^2 {\rm s}^{-1}.
\label{Solomon}
\end{align}
Although this scaling law describes well the rotation rate of MCS (see below), it cannot be applied to evaluate the angular momentum captured by BH accretion, because it neglects the contribution of density fluctuations, which is the dominant one in supersonic turbulence. 

In highly supersonic turbulence, the contribution of net rotation from velocity fluctuations, which leads to Eq.\,(\ref{Solomon}), is negligible at small scales where velocity fluctuations are $\ll \sigma_{v,0}$, and the first contribution from density fluctuations dominates. This leads to a linear scaling, $\langle {\bs j}^2\rangle^{1/2}= (\beta/6)^{1/2} \, \sigma_{v,0} \, R$, where $\beta$ is the exponent of the density correlation function ($\beta=0.61$ in our simulation), because the standard deviation of the center-of-mass offset scales linearly with $R$. In Methods, this result is demonstrated analytically and also confirmed with our numerical simulation (which obeys the observed velocity-size relation). Considering the angular momentum with respect to the position and velocity of a star, the same linear scaling applies, with the standard deviation of the relative velocity between the star and the gas, $\sigma_{v,\rm rel}$, instead of $\sigma_{v,0}$,   
\begin{equation}
\langle \bs j^2 \rangle^{1/2} = (\beta/6)^{1/2} \, \sigma_{v,\rm rel} \, R.
\label{result}
\end{equation}
\begin{figure}[t]
\centering
\includegraphics[width=1.0\textwidth]{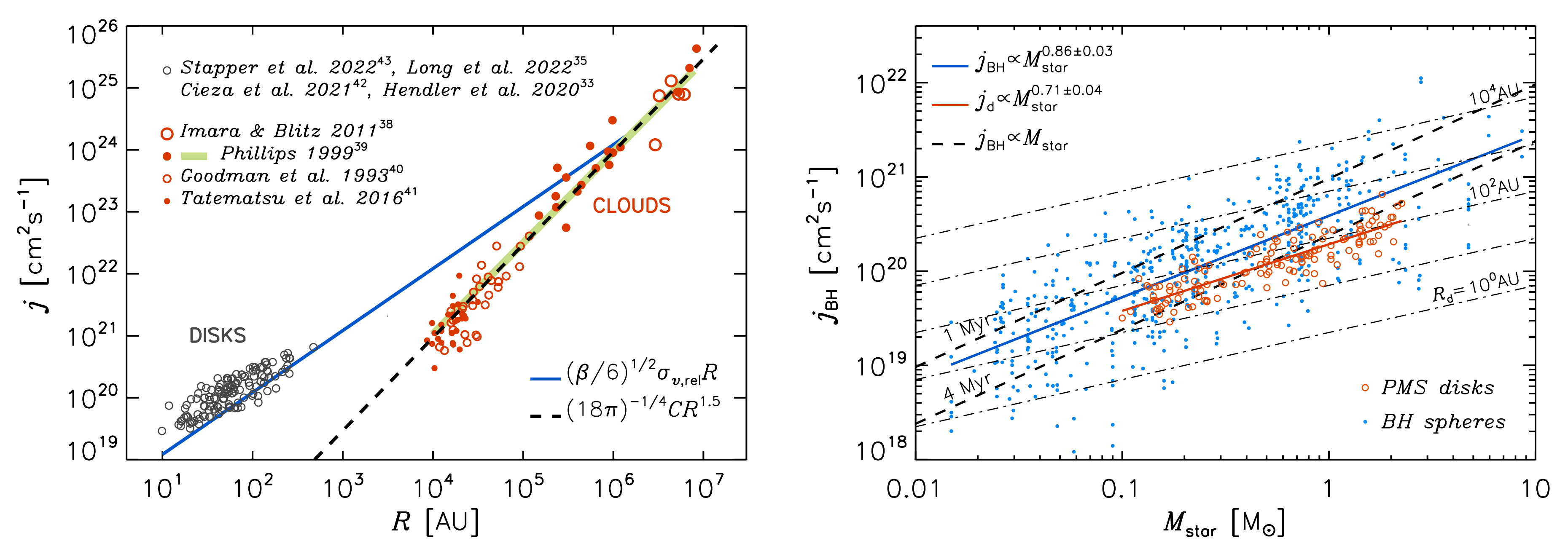}
\caption{Scaling relations of the specific angular momentum. Left: Specific measured angular momentum versus size for individual MCs \citep{Imara+Blitz2011,Phillips99} and cores \citep{Goodman+93,Tatematsu+16}, shown as red circles, and for disks of PMS stars \citep{Hendler+20,Cieza+21,Long+22,Stapper+22}, shown as black circles. A least-squares fit for a large compilation of surveys of clouds and cores \citep{Phillips99} (partially overlapping with the individual clouds shown here) is shown by the solid green line (it almost exactly overlaps with the dashed line between approximately $10^4$ and $10^7$\,AU). The black dashed line is the predicted MC relation from Equation\,(\ref{Solomon}), and the blue solid line the scaling of $\langle j^2\rangle^{1/2}$ as predicted by Equation\,(\ref{result}), using the value of $\sigma_{v,\rm rel}$ from the simulation. Right: Specific gas angular momentum within the BH radius, $j_{\rm BH}$, versus stellar mass, $M_{\rm star}$, measured in 7 snapshots of the simulation, for all sink particles identified as accreting PMS stars (blue dots), with least-squares fit shown by the blue solid line. The two dashed lines correspond to $j_{\rm BH}$ predicted in Equation\,(\ref{j_BH_t_main}), for $t=1$\,Myr (upper line) and 4\,Myr (lower line). The dashed-dotted lines correspond to the disk specific angular momentum, $j_{\rm d}$, for given disk radii, as in Equation\,(\ref{eq_j_d}). The red empty circles give the observational values of $M_{\rm star}$ and $j_{\rm d}$ for PMS stars with resolved disk sizes \citep{Hendler+20,Cieza+21,Long+22,Stapper+22}, with the least-squares fit shown by the red solid line.}
\end{figure}

The left panel of Figure\,1 shows the linear scaling of $\langle \bs j^2  \rangle^{1/2}$ from Equation\,(\ref{result}) with the value of $\sigma_{v,\rm rel}$ taken from the simulation (blue solid line), as well as the steeper scaling from Equation\,(\ref{Solomon}) (black dashed line). The values of $j$ derived in MCs and dense cores \citep{Goodman+93,Phillips99,Imara+Blitz2011,Tatematsu+16}, shown as red circles, are clearly consistent with the predicted scaling, as is a least-squares fit of observational data including some of the objects shown here and others,
$j = 8.7 \times 10^{22} (R/1\,{\rm pc})^{1.47} {\rm cm}^2 {\rm s}^{-1}$ \citep{Phillips99}, shown by the solid green line. The two scaling laws are comparable at the turbulence outer scale, where velocity fluctuations are of the order of $\sigma_{v,0}$. At scales of order $10^2-10^3$\,AU, relevant to BH accretion on PDs, the steeper scaling law of MC measurements based on rotation rate underestimates the general scaling of supersonic turbulence contributions to angular momentum (blue line) by orders of magnitude. On the other hand, the correct prediction of supersonic turbulence contributions leads to $j$ values comparable to those of observed PDs, shown as black circles in the left panel of Figure\,1. The plotted values of $j$ for PDs are derived from the observed disk radius (based on dust continuum observations; see Methods, Section\,\ref{disk_sizes}) and the stellar mass \citep{Hendler+20,Cieza+21,Long+22,Stapper+22}, assuming a simple model of a Keplerian disk (Methods, Section\,\ref{disk}). The disks have $j$ values slightly above the predicted linear scaling. However, as explained in the following, the angular momentum captured by the disks through BH accretion comes from a scale $\sim 4$ times larger than the resulting disk size, where the values of $j$ is larger and more than sufficient to provide the disk angular momentum (imagine the black circles in the left panel of Figure\,1 shifted to the right by a factor of $\sim 4$, bringing them slightly below the blue line).

We now assume that the specific angular momentum of the gas captured by a PMS star moving through the parent cloud is that of the turbulence inertial range at a scale equal to the Bondi-Hoyle radius of the star. We use the following expression for the Bondi-Hoyle radius:
\begin{equation}
R_{\rm BH}=\frac{2 G M_{\rm star}}{c_{\rm s}^2+v_{\rm rel}^2}, 
\label{R_BH}
\end{equation}
which reduces to the Hoyle-Lyttleton radius, $R_{\rm HL}=2 G M_{\rm star}/v_{\rm rel}^2$, in the pressureless case where $c_{\rm s}=0$ \citep{Hoyle+Lyttleton39}, and to the Bondi radius for spherical accretion, $R_{\rm B}=G M_{\rm star}/c_{\rm s}^2$, in the limit of $v_{\rm rel} = c_{\rm s}$ \citep{Bondi52}. The significance of the Bondi-Hoyle radius is that gas streaming at a speed $v_{\rm rel}$ relative to the star, within a minimum distance $\le R_{\rm BH}$, is gravitationally captured by the star ($v_{\rm rel}$ is essentially the star's escape speed at the distance $R_{\rm BH}$). Using $c_{\rm s}^2+v_{\rm rel}^2=\sigma_{v,\rm rel}^2$ in Equation\,(\ref{R_BH}), and setting $R=R_{\rm BH}$ in Equation\,(\ref{result}), the characteristic value of the specific angular momentum of the gas captured by the star is 
\begin{align}
j_{\rm BH} =  4.3\times 10^{20}\, {\rm cm}^2 \,{\rm s}^{-1} \,(\sigma_{v,\rm rel}/2\,{\rm km \,s}^{-1})^{-1}\,(M_{\rm star}/1\,M_{\odot}),
\label{eq_j_BH}
\end{align}
where we have used the numerically derived value of $\beta=0.61$ in Equation\,(\ref{result}) (Methods, Section\,\ref{angular}), and have adopted a normalization of $\sigma_{v,\rm rel}$ comparable to that found in the simulation. Based on a simple model of a Keplerian disk with angular momentum $j_{\rm d}$ (Methods, Section\,\ref{disk}), and setting $j_{\rm d}=j_{\rm BH}$, the characteristic disk radius is
\begin{align}
R_{\rm d} =
            3.6\times 10^2 {\rm AU} \,(\sigma_{v,\rm rel}/2\,{\rm km \,s}^{-1})^{-2}\, (M_{\rm star}/1M_{\odot}),
\label{eq_R_d_BH}
\end{align}
assuming there is no partial cancellation or transport of the angular momentum of the accreting gas, so the actual disk may be somewhat smaller. Interestingly, $R_{\rm d}$ has the same dependence on $\sigma_{v,\rm rel}$ and $M_{\rm star}$ as $R_{\rm BH}$, resulting in a constant ratio of the two quantities,
\begin{align}
R_{\rm BH}/R_{\rm d}=4.1. 
\label{eq_radii_ratio}
\end{align}

Equations\,(\ref{eq_j_BH}) and (\ref{eq_R_d_BH})
depend on the value of $\sigma_{v,\rm rel}$.
In Methods, we derive the time dependence of $\sigma_{v,\rm rel}$ from the velocity-size relation of the interstellar gas (hence the predicted $j_{\rm BH}$ is inversely proportional to the normalization of Larson's velocity-size relation). We show that $\sigma_{v,\rm rel}$ increases with time because the velocity of a star gradually decouples from that of the gas due to the temporal decorrelation of the turbulence. The derived time dependence leads to the following expressions for the total mass gained by BH accretion to the disk from a time $t$ onward,
\begin{equation}
M_{\rm d} = 3.3\times10^{-2} M_\odot \, (t/1 \,{\rm Myr})^{-4}\,(M_{\rm star}/1\,M_\odot)^{2}, 
\label{M_disk_main}
\end{equation}
the mass-averaged $j$ associated with that mass,
\begin{equation}
j_{\rm BH} = 9.6\times10^{20} {\rm cm}^2\,{\rm s}^{-1}\, (t/1 \,{\rm Myr})^{-1}\,(M_{\rm star}/1\,M_\odot), 
\label{j_BH_t_main}
\end{equation}
and the corresponding disk radius,
\begin{equation}
R_{\rm d} = 1.8\times10^3 {\rm AU} \,(t/1 \,{\rm Myr})^{-2}\,(M_{\rm star}/1\,M_\odot). 
\label{R_d_BH_t_main}
\end{equation}
Equations\,(\ref{j_BH_t_main}) and (\ref{R_d_BH_t_main}) should be considered as upper limits, because the angular momentum of the gas captured by an individual star in its trajectory is generally not constant over a $\sim 1$\,Myr timescale, so there must be some partial cancelation in the average of the angular momentum vector. Using tracer particles, Pelkonen et al. (2024) \citep{Pelkonen+24} find a cancelation effect of approximately a factor of three and nearly independent of stellar mass, confirming the results of this paper with their Lagrangian analysis.

These relations should not be used backward in time for $t<1$\,Myr, as explained in Methods. In addition, they should not be interpreted as a strict prediction of the time evolution of the mass, specific angular momentum, and radius of PDs, because we do not specify the status of the preexisting disk at time $t$, nor the processes involved in mixing the infalling gas with the disk. For example, lower $j$ gas infalling at later times may help support the disk accretion onto the central star rather than cause a reduction in the disk size. The purpose of these relations is instead to specifically show that, starting at $\sim 1$\,Myr, in the middle of their Class II phase, {\em PMS stars can still accrete a mass that is in excess of the observed disk masses and carries a large enough angular momentum to explain the observed disk radii.} 

For example, disk masses in Lupus, with an age of $\sim 2$\,Myr, scale with stellar mass as $M_{\rm d}= 7.5\times 10^{-3}\,M_\odot (M_{\rm star}/1\,M_\odot)^{1.7}$, while in Upper Scorpius, with an age of $\sim 4$\,Myr, $M_{\rm d}= 1.9\times 10^{-3}\,M_\odot (M_{\rm star}/1\,M_\odot)^{2.2}$ \citep{Testi+22}, assuming a gas-to-dust ratio of 100.
In Equation\,(\ref{M_disk_main}), the predicted mass captured between 1 and 2\,Myr is approximately four times larger than the disk masses in Lupus. Between 2 and 4\,Myr it is 32 times smaller, but still relevant for the disk mass budget, only a factor of two smaller than the disk masses in the Upper Scorpius region. In addition, BH infall provides a natural explanation for the steep dependence of $M_{\rm d}$ on $M_{\rm star}$, whose origin is otherwise unexplained.  

The dependence of $j_{\rm BH}$ on $M_{\rm star}$ predicted by Equation\,(\ref{j_BH_t_main}) is shown in the right panel of Figure\,1 for $t=1$\,Myr (upper dashed line) and 4\,Myr (lower dashed line). The figure also shows the specific angular momentum of observed PDs, derived from the observed values of the disk radii and stellar masses of PMS stars with resolved disk sizes \citep{Hendler+20,Cieza+21,Long+22,Stapper+22}, the same observational sample as in the left panel of Figure\,1. The predicted values of $j_{\rm BH}$ at times between 1 and 4\,Myr are large enough to account for the observed disk values. Moreover, the observations confirm the prediction that the disk specific angular momentum should increase with the stellar mass, though the derived slope of $0.71$ ($0.95$ using the observational sample from Andrews 2020 \citep{Andrews20}) is a bit smaller than the predicted one of 1.0. However, the fraction of unresolved disks with sizes smaller than $\sim 20$\,AU ($\sim 70$\% of the disks in recent ALMA surveys \citep{Manara+23}) is skewed towards lower stellar masses, so the real mass dependence is indeed expected to be somewhat steeper. Higher resolution surveys, as well as more data for stars above 2\,$M_{\odot}$ \citep{Beltran+deWit16} are needed for more accurate comparisons. In addition, we stress that our scenario of disk formation does not apply to PMS stars that are dynamically ejected from star-forming regions at relatively large velocities during their formation process or immediately afterwards. Thus, the selection and comparative study of such high-velocity PMS stars may shed light on the role of BH infall on disk formation.

We test the theoretical prediction for $j_{\rm BH}$ with our simulation, by measuring the magnitude of the specific angular momentum in spheres around the PMS stars (Methods, Section\,\ref{BH_sim}), which is shown as a function of the stellar mass in the right panel of Figure\,1 (blue dots). The least-squares fit gives $j_{\rm BH} \propto M_{\rm star}^{0.86\pm 0.03}$ (blue solid line) almost identical to the theoretical prediction.
The normalization is a bit lower than in Equation\,(\ref{j_BH_t_main}), considering the median age of 0.92\,Myr of the PMS stars in the simulation, but this is expected because of the slightly larger gas velocity normalization in the simulation relative to the velocity-size relation adopted here (Methods, Section\,\ref{timescale}). The slope is also a bit smaller than the predicted linear relation, likely because the increase of the disk size with stellar mass means that the settling of the infalling gas towards a disk, causing partial $\bs j$ cancellation, is comparatively better resolved for the more massive stars.

\begin{figure}[t]
\centering
\includegraphics[width=0.7\textwidth]{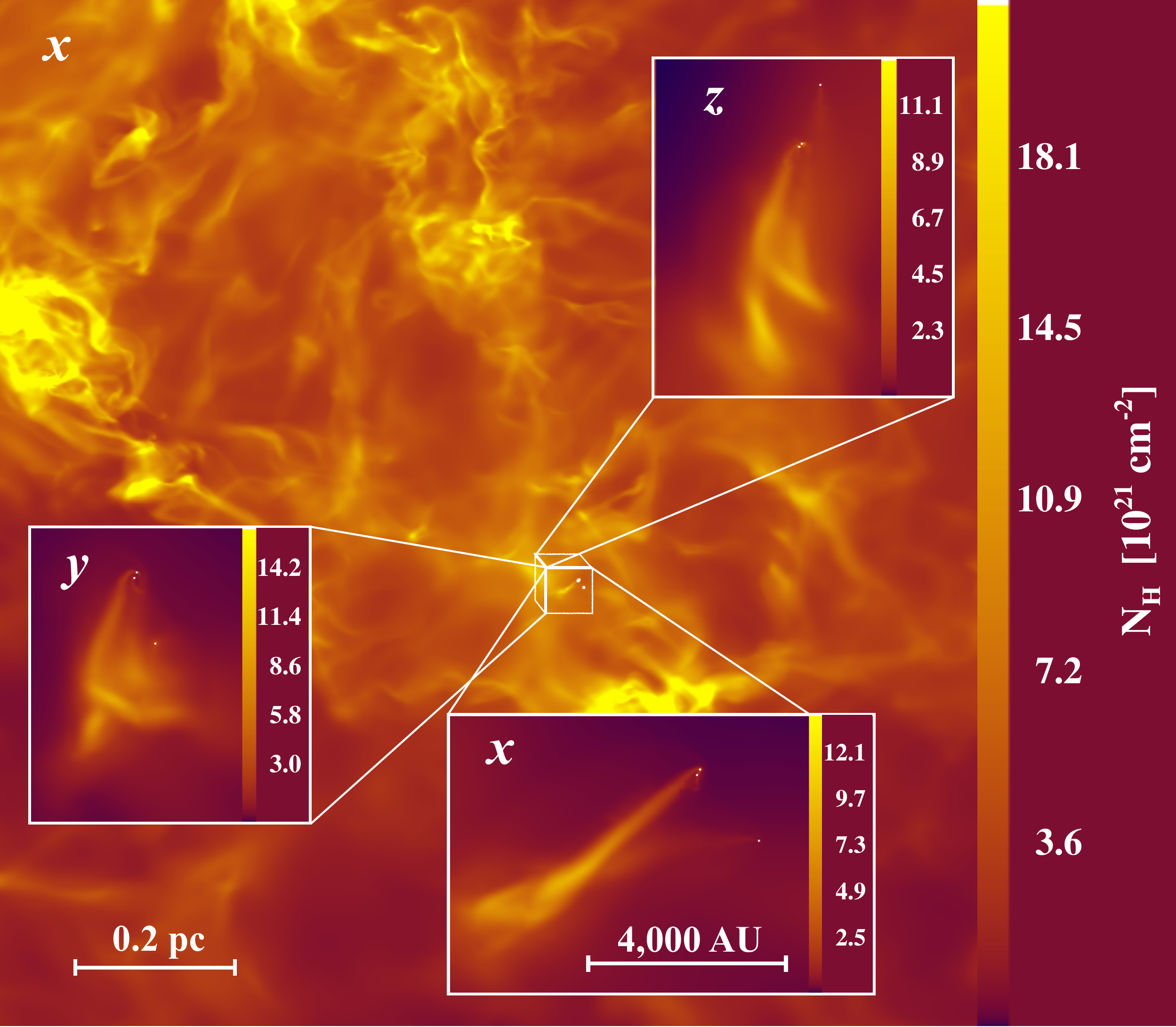}
\caption{Column density images of sub-volumes of the simulation. The entire computational domain is a cube with a side length of 4\,pc. The main image is the projection of a sub-cube with dimensions 1.2\,pc, approximately centered around a PMS bound triple system (indicated by white dots). The three insets show parts of the three orthogonal projections of a 10,000\,AU volume, positioned to capture the BH tails of the triple system. As the stars orbit, their extended BH tails interweave and twist. The Keplerian disks around each star are not visible as they are too small to be resolved in this simulation. The color scale is approximately proportional to the square root of the column density.}
\end{figure}

Besides its significance in terms of mass and angular momentum,
BH infall on PMS stars can strongly affect the evolution of PDs as a consequence of its highly asymmetric nature. Because $\sigma_{v,\rm rel}\gg c_{\rm s}$, the gravitationally captured or deflected head-wind gas that does not collide directly with the disk ($R_{\rm BH}>R_{\rm d}$) shocks onto a wake trailing the star, creating dense filaments (see example in Figure\,2) whose interior parts closer to the star fall back onto the disk. Because of the rather high density of the infalling gas, its effect can be strongly focused on a limited disk region causing appreciable perturbations. On the other hand, because of their low column density, such filaments may escape detection. In a separate work, we demonstrate that dedicated JWST observations can successfully detect them \citep{Pelkonen+24}. 

A general scenario of late-time mass infall onto PDs is consistent with recent discoveries of large-scale flows feeding young disks \citep[e.g.][]{Pineda+20,Alves+20,Huang+21,Grant+21,Valdivia+22,Cacciapuoti+23}, the detection of reflection nebulae around Class II stars \citep{Gupta+23}, and previous numerical studies following the early evolution of disks in realistic large-scale environments \citep[e.g.][]{Kuffmeier+17,Kuffmeier+20,Kuffmeier+21,Kuffmeier+23}.
If further confirmed by future observations, this alternative scenario will compel major revisions of current disk evolution and planet formation models.

\section{Methods}\label{methods}

\subsection{Angular-Momentum Scaling in a Turbulent Flow} \label{angular}

We derive the scaling of the angular momentum within a spherical region of radius, $R$, in a compressible turbulent flow. Without loss of generality, we assume that the region is centered at the origin and, for mathematical convenience, apply a Gaussian filter of size $R$ to evaluate the mass, 
 \begin{equation}
\bs M = \int  \rho \exp\left(-\frac{r^2}{R^2}\right) d^3 r,
\label{massdef}
\end{equation}
and angular momentum,
\begin{equation}
\bs J = \int  (\rho \bs r \times \bs v ) \exp\left(-\frac{r^2}{R^2}\right) d^3 r,
\label{jdef}
\end{equation}
where  $\rho$ and $\bs v$ are density and velocity at $\bs r$. The specific angular momentum is defined as $\bs j= \bs J/ M$, and we aim to calculate the rms of $\bs j$ as a function of $R$. We will assume $ \langle j^2 \rangle =  \langle J^2 \rangle/\langle M^2 \rangle $, which holds at high accuracy, as verified by simulation data. 

From Equation\,(\ref{jdef}), the variance of ${\bs J}$ is calculated as, 
\begin{align}
\langle \bs J^2\rangle =\int d^3 r_1 \int d^3 r_2  \left [  r_{1i }r_{2i}  \langle \rho_1 \rho_2 v_{1j} v_{2j}\rangle  -  r_{1i}r_{2j} \langle \rho_1 \rho_2 v_{1i} v_{2j} \rangle \right]
\exp\left(-\frac{r_1^2 +r_2^2}{R^2}\right),
\end{align}
where the subscripts ``1" and ``2" indicate quantities at two points $\bs r_1$ and $\bs r_2$, respectively. Under the assumption of statistical homogeneity and isotropy, it is straightforward to show that $\langle \rho_1 \rho_2 v_{1i} v_{2j} \rangle =  B^{\rho, v}_{ij}(\bs s) - \frac{1}{2} B_{\rho} S^{\rm dw}_{ij} (\bs s) $, where the density correlation function $B_\rho(\bs s) \equiv \langle \rho_1 \rho_2\rangle $,
the mixed correlation function $B^{\rho, v}_{ij} (\bs s) \equiv \langle \rho_1 \rho_2 v_{1i}v_{1j} \rangle$
and the density-weighted structure function $S^{\rm dw}_{ij}(\bs s)\equiv \langle \rho_1 \rho_2 (v_{2i} -  v_{1i} )(v_{2j} -  v_{1j} ) \rangle/B_{\rho}$, depend on the separation, $\bs s = \bs r_2 -\bs r_1$. Due to the nonlinear and stochastic nature of turbulence, the relative phase of the velocity and density fields appears to be random, leading to a negligible density-velocity correlation in highly supersonic turbulence, as confirmed by our simulation data (shown below) and other numerical studies \citep[e.g.][]{Rabatin+Collins23}. If the density and velocity fields are assumed to be independent, we have $B^{\rho, v}_{ij}  = \langle \rho_1 \rho_2 \rangle \langle v_{1i} v_{1j} \rangle = B_\rho v'^2 \delta_{ij}$ with $v'$ the 1-dimensional rms velocity, suggesting that the longitudinal and transverse correlation functions can be written as $B^{\rho,v}_{ll} = B^{\rho, v}_{nn} =  B_\rho v'^2$ (see Supplementary Figure\,1 and discussion below).
The above equation can then be rewritten as, 
\begin{align}
\langle \bs J^2\rangle = & 2 v'^2 \int d^3 r_1 \int d^3 r_2   B_\rho(\bs s) 
 r_{1k }r_{2k}  
\exp\left(-\frac{r_1^2 +r_2^2}{R^2}\right) +\notag\\
&\frac{1}{2}\int d^3 r_1 \int d^3 r_2  B_\rho(\bs s) \left [    r_{1i}r_{2j} S^{\rm dw}_{ij}(\bs s) -r_{1k }r_{2k}  S^{\rm dw}(\bs s)  \right]
\exp\left(-\frac{r_1^2 +r_2^2}{R^2}\right).
\label{jsquared}
\end{align}
With isotropy, the structure function $S^{\rm dw}_{ij}(\bs s) = S^{\rm dw}_{nn}( s) \delta_{ij} + [S^{\rm dw}_{ll}(s)-S^{\rm dw}_{nn}(s)]s_i s_j/s^2$, where  $S^{\rm dw}_{ll}$ and $S^{\rm dw}_{nn}$ are the longitudinal and the 
transverse components. Also with the assumption of independence between the density and velocity fields, we would have $S^{\rm dw}_{ij}(\bs s) =S_{ij}(\bs s)$. 

The two contributions in Equation\,(\ref{jsquared}) can be intuitively understood as follows. The first term which arises mainly due to density fluctuations represents the offset of the mass center from the geometric center of sphere (see below), while the second term, which depends on the velocity structure function,  originates from the ``imbalance" of the turbulent velocity on the opposite sides of the geometric center, leading to a ``residual" angular momentum. For convenience, we denote the two terms as $\langle \bs J^2\rangle_1$ and $\langle \bs J^2\rangle_2$, respectively. As discussed below, the first term provides the dominant contribution in the highly supersonic regime. 
It can also be shown that for the weakly compressible or incompressible regime, it is the second term that dominates.   

By changing integral variables, $\bs s = \bs r_2 -\bs r_1$ and $\bs t= \bs r_2 +\bs r_1$, the two terms in Equation\,(\ref{jsquared}) can be simplified by carrying out the integration with respect to $d^3t$, yielding, 
\begin{equation}
\langle \bs J^2\rangle_1 =\frac{(2 \pi)^{\frac{3}{2}} R^3 v'^2}{16} \int   B_{\rho} (s) \left( 3R^2 - s^2 \right)\exp\left(-\frac{s^2}{2R^2}\right) d^3 s,
\label{j21}
\end{equation}
and,
\begin{equation}
\langle \bs J^2\rangle_2 =\frac{(2 \pi)^{\frac{3}{2}} R^3}{32} \int   B_{\rho} (s) \left [ s^2 S^{\rm dw}_{nn}(s)   - R^2 S^{\rm dw}(s) \right]\exp\left(-\frac{s^2}{2R^2}\right) d^3 s.
\label{j22}
\end{equation}
A similar calculation for the variance of $M$ leads to, 
\begin{equation}
 \langle M^2 \rangle = \frac{(2\pi)^{\frac{3}{2}} R^3}{8} \int    B_{\rho} (s) \exp\left(-\frac{s^2 }{2R^2}\right)  d^3s.
 \label{m2}
\end{equation}

For the application to interstellar turbulence, we consider highly supersonic turbulence.  
We first verify the assumptions in our formulation using data from a simulation of supersonic MHD turbulence with a sonic Mach number of $10$ (see Section\,\ref{angular_sim}). The left panel of Supplementary Figure\,1 confirms that the density-velocity mixed correlation functions are approximately equal to the density correlation function times $v'^2$, $B^{\rho,v}_{ll}(s) \approx B^{\rho, v}_{nn}(s) \approx  B_\rho(s) v'^2$, as expected from the statistical independence of density and velocity fields, as mentioned above. The right panel shows that the density-weighted velocity structure functions exhibit similar behaviors as the velocity structure functions, $S^{\rm dw}_{ij}(s) \approx S_{ij}(s)$, except for a slightly larger amplitude of the transversal structure function, $S_{nn}$ (filled blue squares), and a slightly steeper slope of the density-weighted transversal structure function, $S^{\rm dw}_{nn}$ (empty red squares). In addition, the left panel shows that the density correlation function (black circles) can be approximated by a power-law function. A least-squares fit for the density correlation function, $B_\rho(s) \propto s^{-\beta}$, gives $\beta=0.61$, which is the value we adopt in our applications. Assuming that $B_{\rho}(s) = c s^{-\beta}$, we may integrate Equation\,(\ref{m2}) to obtain, 
\begin{equation}
 \langle M^2 \rangle = 2^{1-\frac{\beta}{2}} \pi^{\frac{5}{2}} \Gamma\left(\frac{3-\beta}{2}\right) c R^{6-\beta},
 \label{m2f}
\end{equation}
where $\Gamma$ is the Gamma function. 
Using integration by parts for the integral in Equation\,(\ref{j21}), we find that, 
\begin{align}
\frac{\langle \bs J^2\rangle_1}{\langle M^2 \rangle}= \frac{1}{2}\beta v^{\prime2} R^2, 
\label{mainresult_method}
\end{align}
which suggests that the offset of the mass center from the geometric center due to strong density fluctuations contributes an rms angular momentum $\propto R$ in highly supersonic turbulence. 

To evaluate $\langle {\bs J}^2\rangle_2$, we assume inertial-range scaling for the density-weighted structure functions, $S^{\rm dw}_{ll}(s) = c_{l} s^{\gamma}$ and  $S^{\rm dw}_{nn}(s) = c_{n} s^{\gamma}$, where the parameter $c_{l}$, $c_{n}$ and $\gamma$ may be constrained by numerical simulations. Inserting into Eq.\ (\ref{j22}), we find that,  
\begin{align}
\langle \bs J^2\rangle_2= 2^{\frac{\gamma-\beta}{2}-1} \pi^{\frac{5}{2}}  \Gamma \left( \frac{3+\gamma-\beta}{2}\right) c [(1+\gamma-\beta) c_{n} -c_{l}] R^{8+\gamma-\beta}, 
\label{J2result}
\end{align}
so that, 
\begin{equation}
\frac{\langle \bs J^2\rangle_2}{\langle M^2 \rangle}
= 2^{\frac{\gamma}{2}-2}\frac{\Gamma \left( \frac{3+\gamma-\beta}{2}\right) }{\Gamma\left(\frac{3-\beta}{2} \right)}\left[(1+\gamma-\beta) c_{n} -c_{l}\right] R^{2+\gamma}.
\label{j2result}
\end{equation}
Clearly, this contribution depends on velocity scaling, which is expected as it originates from velocity fluctuations within the sphere. The $R^{2+\gamma}$ scaling for $\langle j^2 \rangle_2$ could be derived from a simple dimensional analysis. 
 
In the highly supersonic regime, where $\beta \sim 1$ ($\beta=0.61$ in our simulation), it is straightforward to see that that $\langle \bs J^2\rangle_1$  dominates over $\langle \bs J^2\rangle_2$ at inertial-range scales $R$ much smaller than the integral scale, $L$, of the flow.  This is because at $R \ll L$, 
$S^{\rm dw}_{ll}(R) =  c_l R^\gamma \ll v'^2$ and $S^{\rm dw}_{nn}(R) = c_n R^\gamma \ll v'^2$. Only when $R$ approaches $L$, could the two contributions become comparable. Since $\langle \bs J^2\rangle_1 \gg \langle \bs J^2\rangle_2$ for small scales, we have, 
\begin{align}
\langle \bs j^2\rangle = \frac{1}{2}\beta v^{\prime2} R^2, 
\label{mainresult_m}
\end{align} 
which is found to be in excellent agreement with results from our simulation of highly supersonic turbulence (see Section\,\ref{angular_sim}). Supplementary Figure\,2 shows the scaling of $\langle \bs j^2\rangle^{1/2}$ from both the simulation (blue squares) and the analytical result (red dashed line). Both the slope and the normalization are nearly identical in the two cases, for distances within the limited inertial range of the turbulence in the simulation. Small deviations appear only at large scale, affected by the driving force, and small scale, affected by numerical dissipation. 

We offer a more intuitive derivation of the linear scaling of $\langle \bs j^2  \rangle^{1/2}$ in highly supersonic turbulence, Equation\,(\ref{mainresult_m}), where the role of the mass center offset is more easily appreciated. 
We consider $R$ much smaller than the integral scale of turbulence, so that one may neglect fluctuations of ${\bs v}$ and assume it is constant within the sphere, ${\bs v}={\bs v}_{\rm c}$. Setting ${\bs v}={\bs v}_{\rm c}$ in Equation\,(\ref{jdef}) for $\bs J$ yields
\begin{equation}
\bs J = M \bs r_c \times {\bs v}_{\rm c},
\label{japprox}
\end{equation}
where $M$ is the total mass $M$ (Equation\,(\ref{massdef})) and $ \bs r_c$ is the mass center defined as $\bs r_c = \frac{1}{M}\int  \rho \bs r \exp\left(-\frac{r^2}{R^2}\right) d^3 r$. Due to strong density fluctuations in the highly supersonic turbulence, $\bs r_c$ may deviate significantly from the geometric center of the sphere, leading to considerable angular momentum, as implied by Equation\,(\ref{japprox}). The variance of  $\bs r_c$ may be estimated as, 
\begin{equation}
\langle \bs r_c^2 \rangle  = \frac {1}{\langle M^2 \rangle} 
\int d^3 r_1 \int d^3 r_2  B_{\rho} (\bs r_2 -\bs r_1) \rangle (\bs r_1 \cdot \bs r_2)  \exp\left(-\frac{r_1^2 +r^2_2}{ R^2}\right).
\end{equation}
Using the power-law scaling for the density correlation function $B_{\rho} \propto s^{-\beta}$, we find that $\langle \bs r_c^2 \rangle = \frac{1}{4} \beta R^2$, suggesting that the mass center offset is linear with $R$. 

Further assuming the independence between the velocity and gas density (as in Section\,\ref{angular}) and the randomness of the velocity direction and considering that the rms of ${\bs v}_c$ is essentially the rms turbulent velocity, Equation\,(\ref{japprox}) would produce the same result, Equation\,(\ref{mainresult_m}), derived earlier for the variance of the specific angular momentum. If we adopt the three-dimensional rms turbulent gas velocity $\sigma_{v,0}$ rather than the 1D rms $v'$, 
we have $\langle \bs j^2\rangle = \frac{1}{6}\beta \sigma_{v,0} R^2$. The linear scaling of $\langle \bs j^2\rangle^{1/2}$ with $R$ originates from the linear scaling of the offset distance $r_c$ with $R$, while the contribution from velocity fluctuations inside the sphere is negligible.

\subsection{Angular Momentum Versus Size for Molecular Clouds} \label{Larson}

In the observational studies, the specific angular momentum of MCs is computed as $j = R^2 \Omega$, where $\Omega$ is the cloud's overall angular speed, derived from the gradient of the mean radial velocity of emission line spectra at different cloud positions. Here we show that this definition of $j$ is equivalent to our definition in Equation\,(\ref{jdef}), if the gas density is assumed to be constant, hence the observed scaling of $j$ in MCs can be predicted from the $\langle {\bs J}^2\rangle_2$ term in our formalism in Section \ref{angular}, imposing constant density. 

In our formulation, the overall angular velocity of a cloud may be estimated as (see Pan et al.\ 2016),
\begin{equation}
\bs \Omega = \frac{1}{2V}\int \bs \omega  \exp\left(-\frac{r^2}{R^2}\right) d^3 r,
\label{rotation}
\end{equation}
where $V$ is the effective volume $V= (2\pi)^{3/2} R^3 $ and $\bs \omega$ is the vorticity of the turbulent velocity in the cloud. Using the Gauss theorem, it follows from the above equation that,
\begin{equation}
\bs j = R^2  \bs \Omega = \frac{1}{V}\int (\bs r \times \bs v) \exp\left(-\frac{r^2}{R^2}\right) d^3 r,
\label{rotation2}
\end{equation} 
which is equivalent to Equation\,(\ref{jdef}) assuming constant density.
The observational method is therefore equivalent to setting the density at a constant value in our formalism in Section\,\ref{angular}. In that case, the density correlation function is $B_{\rho} \simeq \rho_0^2$, where $\rho_0$ is the constant density of the flow, hence $\beta=0$. As a result, the first of the two contributions to the variance of the angular momentum in this case is $\langle \bs J^2\rangle_1=0$. In addition, the second term, $\langle \bs J^2\rangle_2$, is simplified by setting $\beta=0$ in Equation\,(\ref{j2result}). To evaluate the velocity structure functions, we make use of Larson's velocity-size relation \citep{Larson81}, $\Delta v(\ell) = C \ell^{\alpha}$, where $\alpha \simeq 0.5$ \citep{Solomon+87} and $C$ is a constant, and assume equipartition between solenoidal and compressive modes in the power spectrum, so that $S_{ll}(s) = S_{nn}(s) = \frac{1}{3} C^2 s^{2\alpha}$. Thus, Equation\,(\ref{j2result}) is further simplified by setting $c_l=c_n=C^2/3$, and we obtain: 
\begin{equation}
\langle \bs j^2  \rangle = \frac{2^\alpha}{3} \pi^{-\frac{1}{2}} \alpha C^2  \Gamma\left(\frac{3}{2} + \alpha\right) R^{2+2\alpha},
\label{rotation3}
\end{equation} 
If $\alpha =0.5$, we have $\langle \bs j^2  \rangle^{1/2}  = (18\pi)^{-\frac{1}{4} }CR^{\frac{3}{2}}$.  Using the relation from Solomon et al. (1987) \citep{Solomon+87}, $\Delta v(\ell) = 0.72 (\ell/1 $ pc $)^{0.5}$ km s$^{-1}$ (where $\Delta v(\ell)$ is an estimate of the three-dimensional velocity dispersion), we find that
\begin{equation}
\langle \bs j^2  \rangle^{1/2}  = 8.3 \times 10^{22} (R/1  {\rm pc})^{3/2} {\rm cm}^2 {\rm s}^{-1}.
\label{j_observations}
\end{equation}

As shown in the left panel of Figure\,1 in the Main text, this predicted scaling is almost identical to the relationship between the specific angular momentum and the size of MCs derived from the observational data. The scaling of the angular momentum of MCs and dense cores is not directly applicable to our problem of estimating the angular momentum of the gas captured by a PMS star along its trajectory. If applied to our problem, it would significantly underestimate the specific angular momentum of the gas captured by the star, as shown by the left panel of Figure\,1. The two scaling laws shown in that figure are comparable at the turbulence outer scale (the computational box size in this case), where $\sigma_{v,0}$ (or the corresponding $\sigma_{v,\rm rel}$) is computed, because at that scale the velocity fluctuations (as well as net rotation) are of the order of $\sigma_{v,0}$. With a larger outer scale, the normalization of the linear scaling law would be larger, further increasing the discrepancy with the MC's scaling law at small scales.

\subsection{Numerical Simulation and $j$ Scaling} \label{angular_sim}

In order to test the analytical results, we use a numerical simulation of randomly driven, supersonic, magnetohydrodynamic (MHD) turbulence, designed to simulate the star-formation process in a turbulent interstellar cloud. The simulation, also used in other recent works \citep{Jorgensen+22,Jensen+23,Kuffmeier+23,Kuffmeier+24,Pelkonen+24}, is equivalent to the \emph{high} reference simulation in Haugb{\o}lle et al. (2018) \citep{Haugbolle+18imf}, except that the numerical resolution (the root grid) is larger by a factor of two (a factor of 8 increase in the number of cells). The reader is referred to that work for details of the numerical methods, which is only briefly summarized here. The simulation solves the MHD equations with the adaptive mesh-refinement (AMR) code Ramses \citep{Teyssier07}, with a root grid of $512^3$ cells and six levels of refinement, corresponding to a smallest cell of size $\Delta x=25$\,AU for the assumed box size of 4\,pc. The total mass, mean density, and mean magnetic field strength are $M_{\rm box}=3000 \rm \,M_\odot$, ${\bar n_{\rm H}}=1897 \, \rm cm^{-3}$, and ${\bar B}=7.2 \, \mu \rm G$, appropriate for a typical star-forming cloud at that scale. The equation of state is assumed to be isothermal, and the boundary conditions are periodic. The turbulence is first driven, without self-gravity, for $\sim 20$ dynamical times, with a random solenoidal acceleration giving an rms sonic Mach number of approximately 10. Keeping the driving force, the simulation is then continued for $\sim 2$\,Myr with self-gravity and sink particles to capture the formation of individual stars, yielding 317 stars with a mass distribution consistent with the observed stellar IMF, and a final star formation efficiency of approximately 8\%. \citep{Salpeter55,Chabrier05}. Feedback from high-mass stars is not included, as only two stars with mass close to $10\,M_{\odot}$ are found towards the end of the simulation, and their structure should probably be inflated due to the high accretion rate, thus preventing a significant UV flux at this early stage. In addition, all the nearby star-forming regions whose disk sizes are reported here are also devoid of high-mass stars.

To test Equation\,(\ref{mainresult_m}), we generate density and velocity snapshots in a uniform grid of $512^3$ cells (using only the root grid of the AMR simulation), and compute the angular momentum, ${\bs J}$, and the mass, $M$, within spherical volumes of radius $R$ with a Gaussian cutoff, consistent with Equations\,(\ref{jdef}) and (\ref{massdef}). We then compute the specific angular momentum, ${\bs j}={\bs J}/M$, for each sphere. We use $16^3$ spheres with centers uniformly distributed in the computational domain, and collect data from six snapshots, so the variance of ${\bs j}$ is computed from an average over 16,384 spherical volumes. The procedure is repeated for 7 values of the cutoff radius, $R=4$, 8, 16, 32, 64, 128, 256$\times \Delta x$. The result is shown by the filled blue squares in Supplementary Figure\,2. The least-squares fit (solid black line in Figure\,4) has both slope and normalization indistinguishable from those predicted by Equation\,(\ref{mainresult_m}) (red dashed line in Figure\,4). 

The same 6 snapshots were also used to test the key assumption of the derivation of Equation\,(\ref{mainresult_m}), that is the independence of density and velocity fields, and to measure the scaling exponent ($\beta=0.61$) of the density correlation function, $B_\rho(s)$ (see Section\,\ref{angular}).

\subsection{Disk Model} \label{disk}

We consider a very simple disk model, for the sole purpose of relating the estimated $j_{\rm BH}$ to a characteristic disk size. The disk is assumed to have a power-law column density profile with exponent $n$ and to be truncated at an outer radius $R_{\rm d}$. For $n<2$, the inner radius, $R_{\rm i}$, is irrelevant for the normalization to the total mass, as long as $R_{\rm i}\ll R_{\rm d}$, and could also be zero, and we can write the radial dependence of the column density as
\begin{equation}
\Sigma_{\rm d}(R)=\frac{(2-n)M_{\rm d}}{2 \pi R_{\rm d}^2}\left(\frac{R}{R_{\rm d}}\right)^{-n}
\label{eq_Sigma}
\end{equation}
where $M_{\rm d}$ is the total disk mass. Assuming the disk has a Keplerian velocity profile, its mass-averaged specific angular momentum is given by
\begin{align}
j_{\rm d} &= \frac{4-2n}{5-2n}G^{1/2}M_{\rm star}^{1/2}R_{\rm d}^{1/2} \nonumber \\
           &= 2.25\times 10^{19} {\rm cm}^2 {\rm s}^{-1}  \left(\frac{M_{\rm star}}{1M_{\odot}}\right)^{1/2} \left(\frac{R_{\rm d}}{1{\rm AU}}\right)^{1/2}
\label{eq_j_d}
\end{align}
where the second equality assumes $n=3/2$, a typical value in disk models, such as for the minimum-mass solar nebula \citep{Weidenschilling77}.

\subsection{Observed Disk Sizes} \label{disk_sizes}

The stellar mass and disk size values (and the corresponding disk $j$ values) shown in the two plots of Figure\,1 are all measured from PMS stars (i.e. Class II sources). They are taken from four different publications \citep{Hendler+20,Cieza+21,Long+22,Stapper+22}, avoiding duplication when the same objects appear in different samples. For the sake of statistics, the disk radii are measured from dust continuum observations, $R_{\rm mm}$, rather than gas observations, $R_{\rm CO}$, as the latter are available only for a smaller number of disks. The values of $R_{\rm mm}$ are taken from the published tables, and are in all cases defined as the radius containing 90\% of the sub-millimeter flux of the disk. It is well established that PDs appear to be a few times larger when traced by gas rather than dust, which is often interpreted as evidence of growth and radial settling of dust grains. For example, for the disks from Long et al. (2022) \citep{Long+22} that have both types of sizes, $R_{\rm CO}/R_{\rm mm}=2.9\pm1.2$, on average. However, the ratio does not depend on stellar mass, hence adopting $R_{\rm CO}$ instead of $R_{\rm mm}$ in our work would not change our conclusions on the mass dependence. The derived $j$ depends on the $R_{\rm d}^{1/2}$, as shown by Equation\,(\ref{eq_j_d}), so it would increase on average by only 70\%, without affecting our conclusion that BH accretion provides a large enough angular momentum to explain the observed disk sizes. To illustrate the comparison of our analytical and numerical results with the observations based only on the gas radii, $R_{\rm CO}$, we have made a second version of the right panel of Figure\,1, where the values of $j$ for the disks are based on their $R_{\rm CO}$ (Supplementary Figure\,3). The dependence of the disk $j$ on the stellar mass is essentially the same as for $R_{\rm mm}$ (the slope of the least-squares fit is the same within the $1\sigma$ uncertainty), and the $j$ values are mostly within the analytical prediction for ages between 1 and 4\,Myr, with their line of best fit very close to that for the numerical data points.

\subsection{Angular Momentum of Bondi-Hoyle Infall in the Simulation} \label{BH_sim}

The simulation can be used also to measure $j_{\rm BH}$ relative to the position and velocity of PMS stars, and to compare the result with the prediction of Equation\,(\ref{eq_j_BH}). Although new stars are continuously formed in the simulation, towards the end of the run a significant fraction of them have ages in the approximate range 0.5-2.0\,Myr, old enough to be representative of Class II PMS stars. However, because the time it takes to assemble a star may vary from star to star, and can be relatively long ($\sim 1$\,Myr) for massive stars \citep{Padoan+20massive,Pelkonen+21}, we select PMS stars based on the local gas density, rather than the stellar age, which better reflects the observational SED classification as well. For that purpose, we use seven snapshots at regular time intervals covering the last 0.75\,Myr of the simulation, yielding a total of 1,629 stellar positions.  

Averaging over all seven snapshots, we find that
the gas density sampled in spheres of radius $\sim 400$\,AU (of the order of the size of the largest observed disks) centered around the stars, $P_{\rm st}(n)$, has a clear bimodal distribution, shown by the shaded blue histogram in the left panel of Supplementary Figure\,4. The red unshaded histogram shows the overall gas density distribution sampled uniformly in the whole volume at the same resolution, $P_{\rm V}(n)$.  The lower-density peak of $P_{\rm st}(n)$ follows approximately the overall distribution, though shifted to slightly larger density and with a shape a bit skewed to the right, while the higher-density peak has a maximum at a density larger by four orders of magnitude. The stars at such high local density are still embedded in their native dense gas and are generally increasing their mass at a high rate. The black circles in the left panel of Supplementary Figure\,4 show the gas infall rate on the sink particles in the simulation (with values shown on the right $y$ axis). It strongly correlates with the local gas density, and the infall rates of stars below $\sim 10^5$\,cm$^{-3}$ are of the same order of magnitude as the observed accretion rates of young PMS stars. 

We have verified through visualizations that most of these stars at lower densities are accompanied by gas structures with morphology and kinematics consistent with BH trails, and their predicted BH infall rate is also of the order of the measured infall rate in the simulation, as shown in the right panel of Supplementary Figure\,4. We have also verified that, on average, the local density is inversely correlated with the stellar age. The stars at low density have clearly decoupled from their native dense cores and tend to sample the random density and velocity fields of the parent cloud at larger scales. Based on these results, we choose a fixed critical density value, $n_{\rm H,cr}$, to select the sink particles representative of Class II PMS stars in the BH phase, $n_{\rm H,cr}=5\times 10^4$\,cm$^{-3}$, shown by the vertical dashed line in Figure\,6. This selection yields 961 PMS stars, out of the total 1,629 stars found in the seven snapshots.

We measure the mean mass, gas velocity and angular momentum within spheres of different radii centered on the position of each star. The density and velocity fields of the whole 4\,pc box are first extracted into a uniform grid of 1,024$^3$ cells, so the cell size is $dx=0.004$\,pc or 780\,AU. The spheres have radii of 2, 4, 8, 16, 32, 64, 128, and 256\,$dx$, with the two smallest and the largest expected to be outside the inertial range of the simulation. The gas mass and the velocity and angular momentum components are averaged within each sphere with a Gaussian cutoff, as in Equation\,(\ref{jdef}). 

We estimate the angular momentum of the infalling gas by computing $j$ within a sphere of radius equal to $R_{\rm BH}$, so we need to compute $R_{\rm BH}$ as defined in Equation\,(\ref{R_BH}). The mass and velocity of each star are known, as well as the isothermal sound speed in the simulation, $c_{\rm s}=0.18\times 10^5$\,cm\,s$^{-1}$. To compute $v_{\rm rel}$, the difference between the star velocity and the gas velocity, we use the gas mean velocity measured within a sphere centered on the star. Because the flow is turbulent, the mean gas velocity may vary when measured at different scales, so we should compute it at a scale larger than $R_{\rm BH}$, but not too large. We settle on a radius of 8\,$dx$=6,250\,AU, which is significantly larger than the largest values of $R_{\rm BH}$ and the typical PD sizes. For the sink particles, we find values of $R_{\rm BH}$ in the approximate range 1-3,000\,AU. We cannot measure $j$ directly from the simulation at such scales, because they are well within the numerical dissipation range. Instead, we measure it within spheres of radius of 8\,$dx$=6,250\,AU, as for the estimate of $v_{\rm rel}$, because this is the smallest size we can consider without being significantly affected by numerical dissipation. We then extrapolate the value of $j$ at $R=R_{\rm BH}$, using the linear $j$ scaling established earlier both analytically and numerically (see Supplementary Figure\,2). 

The results are shown in the right panel of Figure\,1 as a function of the stellar mass for all the selected PMS stars in the simulation (blue dots). The least-squares fit has a slope of 0.86 (solid blue line), a bit shallower than the one predicted in Equation\,(\ref{j_BH_t_main}). The plot shows a significant scatter at any given value of $M_{\rm star}$, as expected from Equation\,(\ref{eq_j_BH}) due to the dependence on $\sigma_{v,\rm rel}$: for an individual star at a given time, $j_{\rm BH}$ depends on the local value of $v_{\rm rel}$, which is a random variable. The scatter and the average value of $j$ in PDs are expected to be somewhat smaller than those of individual BH spheres computed here, due to partial cancelation in the vector sums of $\bs J$ from different BH spheres. The right panel of Figure\,1 shows that observed disks indeed have a slightly reduced mean angular momentum and a smaller scatter than our individual BH spheres, as expected.

\subsection{Time Dependence of PMS Disk Formation by Bondi-Hoyle Infall} \label{timescale}

Whether or not BH infall is so dominant to completely restructure PDs during the Class II phase, that is on a timescale of $\sim 2$\,Myr, depends on the stellar mass and on the time evolution of the relative velocity between the stars and the turbulent gas, $v_{\rm rel}$. The BH infall rate, defined as a mass flux through a surface of area $\pi R_{\rm BH}^2$ with gas velocity $v=(c_{\rm s}^2+v_{\rm rel}^2)^{1/2}$, gas number density $n_{\rm H}$, and $R_{\rm BH}$ given by Equation\,(\ref{R_BH}) is:
\begin{equation}
\dot{M}_{\rm BH}=\frac{4\,\pi\, m_{\rm H}\, n_{\rm H}\, G^2 M^2_{\rm star}}{(c_{\rm s}^2+v_{\rm rel}^2)^{3/2}}.
\label{Mdot_BH}
\end{equation}
Vorticity corrections to this BH accretion rate \citep[e.g.][]{Krumholz+05BH} are not significant in the context of low-mass star formation \citep{Krumholz+06BH}.  
When stars are fully decoupled from the parent gas, their velocities are no longer correlated to the gas velocity, and we can assume that the rms velocities satisfy the relation $\sigma^2_{v,\rm rel}= \sigma^2_{v,0}+\sigma^2_{v,\rm s}$, where $\sigma_{v,0}$ is the gas rms velocity at some large scale, and $\sigma_{v,\rm s}$ is the rms velocity of the stars. However, when a star is formed, its velocity is comparable to that of the nearby gas (except for stars accelerated by dynamical interactions) and then gradually decouples from the gas velocity because of the temporal decorrelation of the turbulence. A turbulent eddy of size $R$, with rms velocity $\sigma_{v}(R)$, has a turnover time $\tau(R)\sim R/\sigma_{v}(R)$. After that time, the star velocity is decorrelated from the gas velocity at the scale $R$, but remains coupled to the turbulent velocity at larger scales. Thus, the relevant $\sigma_{v}(R)$ for BH infall increases with time, as the star velocity decouples from the gas velocity at increasingly larger scales. The star is also accelerated in the local gravitational potential. Neglecting stars that achieve significant acceleration by close encounters with other stars (like disrupted binaries), we assume that stars achieve an rms velocity of the order of the virial velocity at scale $R$, in approximately a dynamical time. According to the two Larson's relations, $\sigma_{v}(R)$ is also of the order of the virial velocity at the scale $R$, so we adopt a simple approximation where $\sigma_{v,\rm s}=\sigma_{v}$, meaning that both rms velocities grow in time at the same rate, so $\sigma_{v,\rm rel}=\sqrt{2}\, \sigma_{v}(R(t))$, where $R(t)=R(\tau)$, that is, we identify the time with the eddy turnover time. Using Larson's velocity-size and density-size relations from Solomon et al. (1987) \citep{Solomon+87}, 
\begin{equation}
\sigma_{v} = 0.72 \,{\rm km\,s}^{-1}\,(R/1 \,{\rm pc})^{1/2}, 
\label{larson_v}
\end{equation}
\begin{equation}
n_{\rm H} = 5.2\times10^3 \,{\rm cm}^{-3}\,(R/1 \,{\rm pc})^{-1},  
\label{larson_d}
\end{equation}
the time dependence of the relative velocity and density is given by
\begin{equation}
\sigma_{v,\rm rel} = 0.75 \,{\rm km\,s}^{-1}\,(t/1 \,{\rm Myr}), 
\label{v_time}
\end{equation}
\begin{equation}
n_{\rm H} = 9.6\times10^3 \,{\rm cm}^{-3}\,(t/1 \,{\rm Myr})^{-2}.  
\label{n_time}
\end{equation}

These relations are hard to test in the simulation, due to the expected scatter, the existence of stars with significant dynamical kicks from close encounters, and the broad range in the extent of the embedded phase preceding the BH phase (or Class II), meaning that the time zero of the decoupling corresponds to a different age for different stars \citep{Padoan+20massive,Pelkonen+21}. However, the simulation shows clear evidence of the predicted trends, as illustrated in Supplementary Figure\,5. The figure shows $v_{\rm rel}$ (upper panel) and $n_{\rm H}$ (lower panel) as a function of age (blue dots) for the stars found in the simulation in the same six snapshots used to compute $j_{\rm BH}$. Binaries are not included (35\% of the stars) due to their more complex dynamical evolution, nor stars with accretion rate $> 5\times 10^{-6}\,M_\odot$yr$^{-1}$ (8\% of single stars) because they are still deeply embedded and far from reaching their final mass and starting to decouple from the gas. In the upper panel, we compute the relative velocity rms, $\sigma_{v,\rm rel}$, in age bins (red squares), and fit the result for $t>0.5$\,Myr, which gives $\sigma_{v,\rm rel}\sim t^{0.8\pm0.1}$ (solid red line). The predicted time dependence from Equation\,(\ref{v_time}), with the velocity normalization increased by a factor 1.6 to match the rms velocity in the simulation, is shown by the black dashed line. The lower panel shows the median of $n_{\rm H}$ in age bins (red squares), a least-squares fit for $t>0.5$\,Myr giving $n_{\rm H}\sim t^{-2.3\pm0.4}$ (red solid line), and the prediction from Equation\,(\ref{n_time}) (dashed black line), also based on the renormalized velocity dispersion.  

Using Equations\,(\ref{v_time}) and (\ref{n_time}), Equation\,(\ref{Mdot_BH}) becomes  
\begin{equation}
\dot{M}_{\rm BH} \approx 1.3\times 10^{-7} M_\odot {\rm yr}^{-1} (t/1 \,{\rm Myr})^{-5}(M_{\rm star}/1M_{\odot})^2. 
\label{Mdot_BH_larson_tau}
\end{equation}
This equation should not be used for $t$ significantly shorter than 1\,Myr, because i) $\sigma_{v,\rm rel}$ would become comparable to $c_{\rm s}\sim 0.2$\,km\,s$^{-1}$, which was neglected; ii) at $t=0.5$\,Myr, $n_{\rm H} = 3.8\times 10^4$\,cm$^{-3}$, almost the same as the threshold density, $n_{\rm H,cr}=5\times 10^4$\,cm$^{-3}$\,cm$^{-3}$ that we have identified as the transition density into the BH phase or Class II in the simulation (see Section\,\ref{BH_sim}); iii) Figure\,7 shows that the scatter is dominant for $t<0.5$\,Myr; iv) we are not concerned with times significantly smaller than 1\,Myr because we are interested in the possibility of forming PDs during the Class II phase. 

We now can estimate the mass gained by PDs through BH infall from the time $t$ onward, $M_{\rm d}=\int_t^{\infty}{\dot{M}_{\rm BH}}dt$. Using Equation\,(\ref{Mdot_BH_larson_tau}), we find
\begin{equation}
M_{\rm d} = 3.3\times10^{-2} M_\odot \, (t/1 \,{\rm Myr})^{-4}\,(M_{\rm star}/1\,M_\odot)^2. 
\label{M_disk}
\end{equation}

From Equations\,(\ref{v_time}) and (\ref{eq_j_BH}), 
we can also derive the mass-averaged value of $j_{\rm BH}$ accumulated from the time $t$ onward, $j_{\rm BH}=\int_{t}^{\infty}{j_{\rm BH} \dot{M}_{\rm BH}}dt\,/\int_{t}^{\infty}{\dot{M}_{\rm BH}}dt$,
\begin{equation}
j_{\rm BH} = 9.6\times10^{20} {\rm cm}^2\,{\rm s}^{-1} (t/1 \,{\rm Myr})^{-1}\,(M_{\rm star}/1\,M_\odot), 
\label{j_BH_t}
\end{equation}
and the corresponding disk radius,
\begin{equation}
R_{\rm d} = 2.6\times10^3 {\rm AU} (t/1 \,{\rm Myr})^{-2}(M_{\rm star}/1\,M_\odot), 
\label{R_d_BH_t}
\end{equation}
assuming there is no partial cancellation of angular momentum. Because $\bs J$ of the gas captured by an individual star in its trajectory is generally not constant over a $\sim 1$\,Myr timescale, there must be some partial cancellation, hence Equations\,(\ref{j_BH_t}) and (\ref{R_d_BH_t}) should be considered as upper limits to the actual PD values.   \\

\newpage

\section{Supplementary Figures}

\renewcommand\thefigure{\arabic{figure}}
\setcounter{figure}{0}

\begin{figure}[h]
\centering
\includegraphics[width=0.99\textwidth]{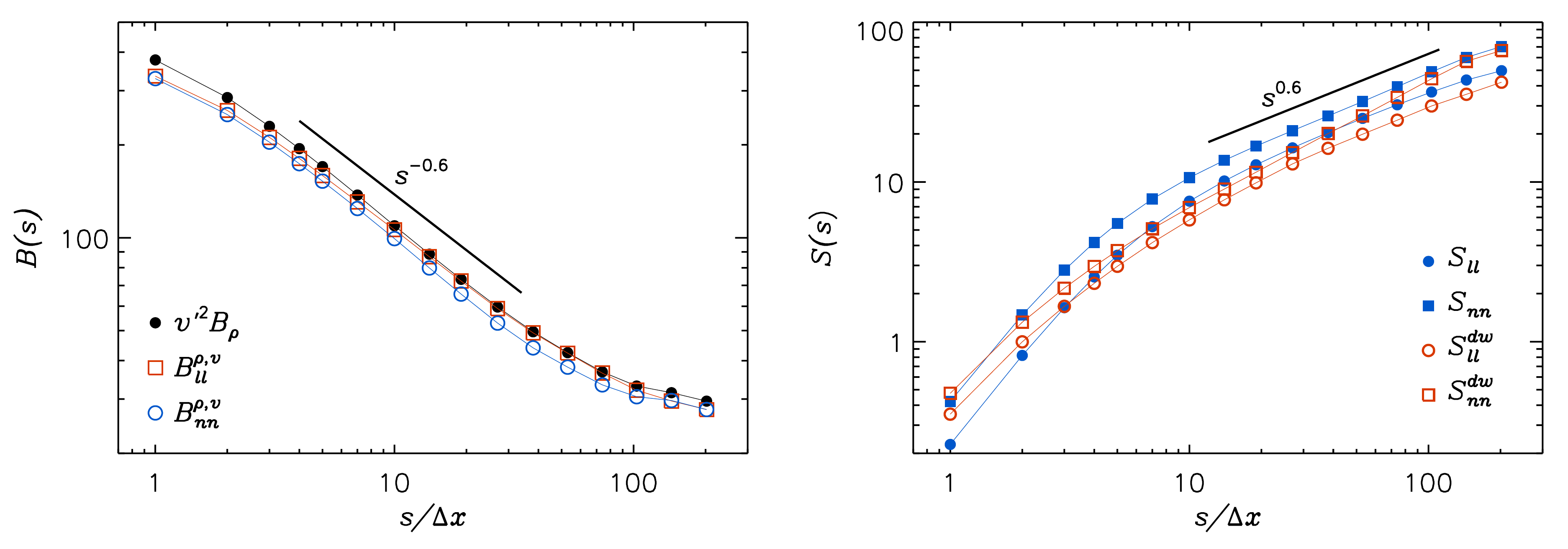}
\caption{Correlation and structure function scalings.
Left: Longitudinal (red squares) and transverse (blue circles) components of the mixed correlation function $B^{\rho, v}_{ij} = \langle \rho_1 \rho_2 v_{1i} v_{2j}\rangle$. Under the assumption of independence between $\rho$ and ${\bs v}$, both are expected to equal $v'^2B_\rho$ (black circles). The density correlation function, $B_\rho$, exhibits a power-law scaling with an exponent of 0.61 in the inertial range. Right: Longitudinal (circles) and transverse (squares) components the velocity structure tensor with (red) and without (blue) density weighting. The structure functions show similar scalings, but the amplitudes of the density-weighted ones are slightly smaller.   
}
\label{correlations}
\end{figure}  


%
\begin{figure}[h]
\centering
\includegraphics[width=0.6\textwidth]{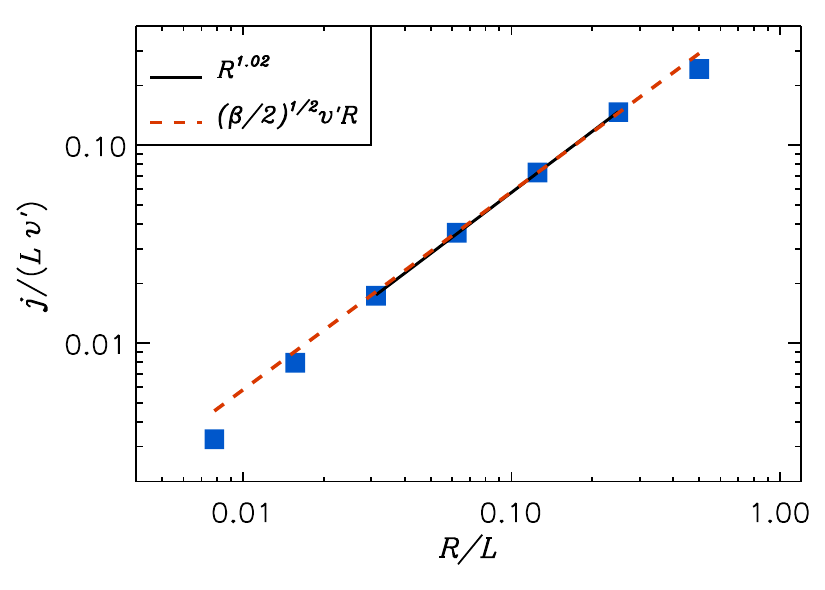}
\caption{Scaling of the variance of the specific angular momentum, $\langle j^2\rangle^{1/2}$, as a function of the size $R$ in units of half the box size, $L$. The blue square symbols and the connected blue lines are the results averaged over four snapshots of the simulation. The red dashed line is the linear prediction from Equation\,(21).
}
\label{j_scaling}
\end{figure} 

\clearpage

\begin{figure}[h]
\centering
\includegraphics[width=0.7\textwidth]
{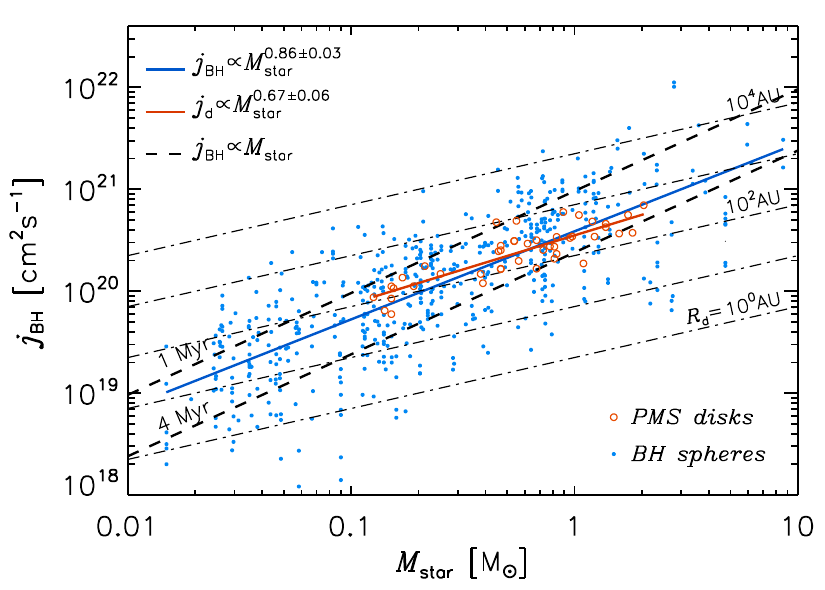}
\caption{Scaling relations of the specific angular momentum. 
The specific gas angular momentum within the BH radius, $j_{\rm BH}$, versus stellar mass, $M_{\rm star}$, measured in 7 snapshots of the simulation, for all sink particles identified as accreting PMS stars, is shown by the blue dots, with the least-squares fit shown by the blue solid line. The two dashed lines correspond to $j_{\rm BH}$ predicted in Equation\,(8), for $t=1$\,Myr (upper line) and 4\,Myr (lower line). The dashed-dotted lines correspond to the disk specific angular momentum, $j_{\rm d}$, for given disk radii, as in Equation\,(29). The red empty circles give the observational values of $M_{\rm star}$ and $j_{\rm d}$ for PMS stars with resolved disk sizes [33, 35, 42, 43], with the least-squares fit shown by the red solid line. This figure is the same as the right panel of Figure\,1, but with the $j$ values of the observed PMS disks derived from the size of the gas disk, $R_{\rm d}=R_{\rm CO}$, instead of the dust disk.
}
\label{j_scaling_GAS}
\end{figure} 

\clearpage

\begin{figure}[h]
\includegraphics[width=0.99\textwidth]{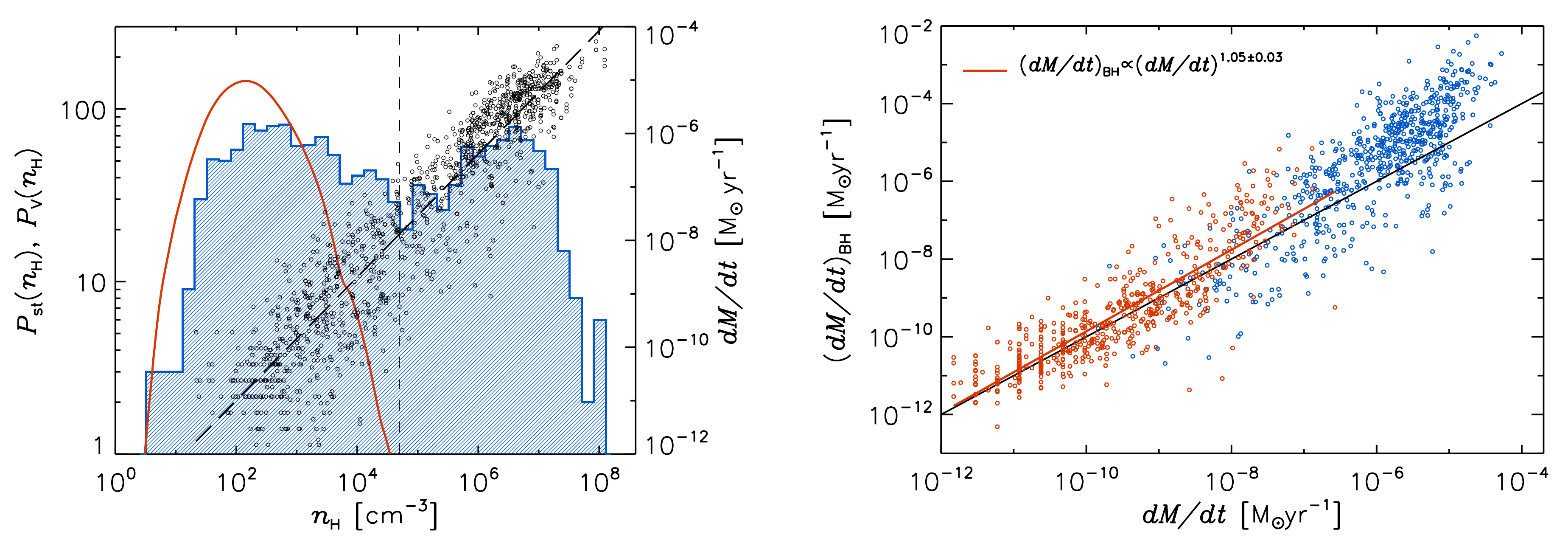}
\caption{Infall rates and local gas density around the stars in the simulation. Left: Probability distribution of density sampled in spheres of radius $\sim 403$\,au centered around the stars, $P_{\rm st}(n_{\rm H})$ (blue shaded histogram), and sampled uniformly in the whole volume at the same resolution, $P_{\rm V}(n_{\rm H})$ (red unshaded histogram), using all seven time snapshots of the simulation. The black points give the infall rate on the stars from the simulation, averaged over a period of 5,000\,yr, with values shown in the right $y$ axis and a linear least-squares fit shown by the long dashed line. The vertical dashed line corresponds to the critical density, $n_{\rm H,cr}=5\times 10^4$\,cm$^{-3}$, used to select the stars representative of Class II objects in the BH phase. Right: The predicted BH infall rate versus the infall rate of the stars in the simulation (as in the left plot). The red symbols are the Class II stars based on the critical density criterion, and the red line is the least-squares fit to the red symbols. The black line is the one-to-one relation.       
}
\label{rho_pdf}
\end{figure} 

\clearpage

\begin{figure}[h]
\centering
\includegraphics[width=0.7\textwidth]{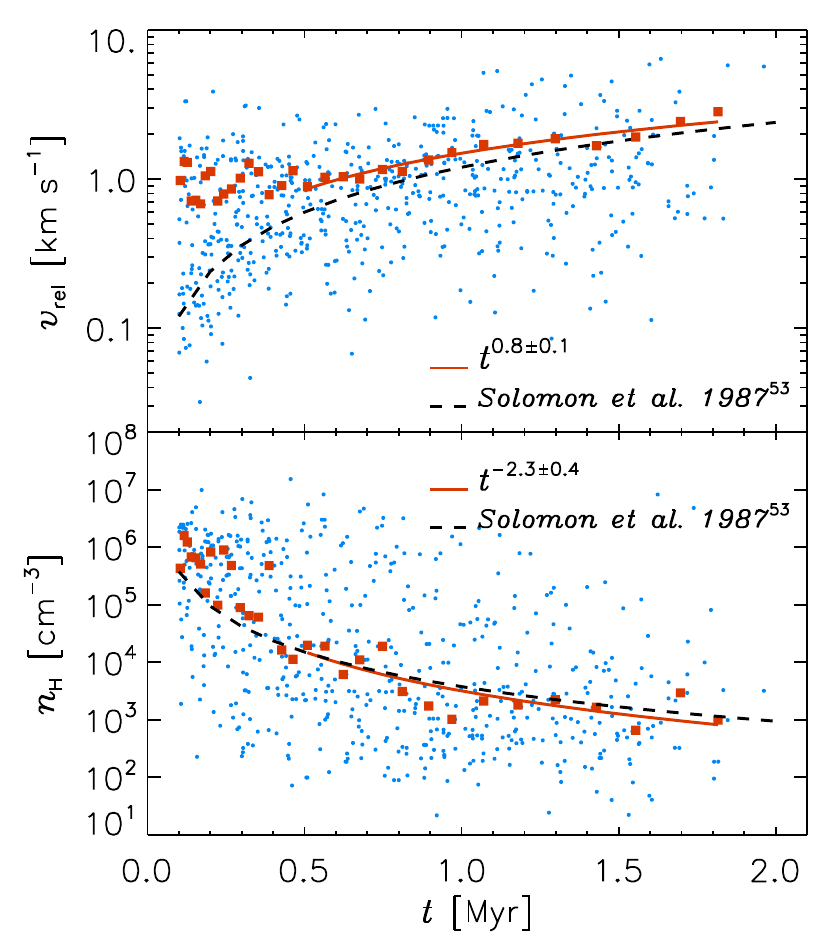}
\caption{Time evolution of the star-gas decoupling. Gas-star relative velocity (upper panel) and gas number density (lower panel) versus star age at the same star positions as in Supplementary Figure\,4, but excluding binaries and stars with accretion rate $> 5\times 10^{-6}\,M_\odot$yr$^{-1}$ (blue dots). Red squares are rms values (upper panel) and median values (lower panel) computed in age bins, with their least-squares fit, for $t>0.5$\,Myr, shown by the red solid lines. The dashed lines are the predictions from Equations\,(21) and (22), with the velocity normalization increased by a factor 1.6, to match the rms velocity in the simulation.    
}
\label{t_vrel}
\end{figure} 

\clearpage

\newpage

\backmatter

\bmhead{Carbon Footprint}
The total cost of the simulation used in this work (and in 7 prior publications where the CO$_2$e was not stated) was approximately 50M core hours (Xeon E5-2680v3), corresponding to 847MWh and a carbon footprint of 121t CO$_2$e (calculated using green-algorithms.org v2.2 \citep{Lannelongue+21_green}).  

\bmhead{Data Availability}
Supplemental material can be obtained from a dedicated public URL (http://www.erda.dk/vgrid/BH-Accretion-of-Disks). This includes tabulated data to reproduce the plots in the paper, and movies of the time evolution of BH trails for some of the stars in the simulation. Raw and post-processed data from the simulation are available upon request (email requests to Paolo Padoan, ppadoan@icc.ub.edu).

\bmhead{Code Availability}
The MHD and gravity solvers used in this study are closely related to the public RAMSES version available at \url{https://bitbucket.org/rteyssie/ramses}. The numerical methods, developed in Copenhagen, relevant for star formation are described in Haugb{\o}lle et al. (2018) \citep{Haugbolle+18imf}.

\bmhead{Acknowledgments}
We thank the anonymous referees for helpful comments.
P.P. acknowledges support from the grants PID2020-115892GB-I00 and CEX2019-000918-M funded by MCIN/AEI/10.13039/501100011033, and by the US National Science Foundation under Grant AST 2408023. L.P. acknowledges support by the NSFC under grants No.\ 11973098 and No.\ 12373072. T.H. acknowledges support from the Independent Research Fund Denmark through grant No. DFF 802100350B. VM.P. acknowledges financial support by the grants PID2020-115892GB-I00 and CEX2019-000918-M funded by MCIN/AEI/10.13039/501100011033, and by the European Research Council via the ERC Synergy Grant ``ECOGAL" (project ID 855130). The Tycho supercomputer hosted at the SCIENCE HPC center at the University of Copenhagen was used for carrying out the postprocessing, analysis, and long-term storage of the results.

\bmhead{Author Contributions Statement} 
P.P. and VM.P. conceived, designed, and carried out the data-analysis study, and L.P. the analytical modeling. T.H. optimized the code and ran the simulation. {\AA}.N. contributed to the code optimization and conceived with P.P. the astrophysical scenario. P.P. and L.P. co-wrote the paper and all other authors contributed edits to the manuscript.

\bmhead{Competing Interests Statement} 
The authors declare no competing interests. \\

\bibliographystyle{sn-nature}

\begin{thebibliography}{10}
\expandafter\ifx\csname url\endcsname\relax
  \def\url#1{\burl{#1}}\fi
\expandafter\ifx\csname urlprefix\endcsname\relax\def\urlprefix{URL }\fi
\providecommand{\bibinfo}[2]{#2}
\providecommand{\eprint}[2][]{\url{#2}}
\providecommand{\doi}[1]{\url{https://doi.org/#1}}
\bibcommenthead

\bibitem{Lesur+23}
\bibinfo{author}{{Lesur}, G.} \emph{et~al.}
\newblock \bibinfo{editor}{{Inutsuka}, S.}, \bibinfo{editor}{{Aikawa}, Y.}, \bibinfo{editor}{{Muto}, T.}, \bibinfo{editor}{{Tomida}, K.} \& \bibinfo{editor}{{Tamura}, M.} (eds) \emph{\bibinfo{title}{{Hydro-, Magnetohydro-, and Dust-Gas Dynamics of Protoplanetary Disks}}}.
\newblock (eds \bibinfo{editor}{{Inutsuka}, S.}, \bibinfo{editor}{{Aikawa}, Y.}, \bibinfo{editor}{{Muto}, T.}, \bibinfo{editor}{{Tomida}, K.} \& \bibinfo{editor}{{Tamura}, M.}) \emph{\bibinfo{booktitle}{Astronomical Society of the Pacific Conference Series}}, Vol. \bibinfo{volume}{534} of \emph{\bibinfo{series}{Astronomical Society of the Pacific Conference Series}}, \bibinfo{pages}{465} (\bibinfo{year}{2023}).

\bibitem{Drazkowska+23}
\bibinfo{author}{{Dr{a}{\.z}kowska}, J.} \emph{et~al.}
\newblock \bibinfo{editor}{{Inutsuka}, S.}, \bibinfo{editor}{{Aikawa}, Y.}, \bibinfo{editor}{{Muto}, T.}, \bibinfo{editor}{{Tomida}, K.} \& \bibinfo{editor}{{Tamura}, M.} (eds) \emph{\bibinfo{title}{{Planet Formation Theory in the Era of ALMA and Kepler: from Pebbles to Exoplanets}}}.
\newblock (eds \bibinfo{editor}{{Inutsuka}, S.}, \bibinfo{editor}{{Aikawa}, Y.}, \bibinfo{editor}{{Muto}, T.}, \bibinfo{editor}{{Tomida}, K.} \& \bibinfo{editor}{{Tamura}, M.}) \emph{\bibinfo{booktitle}{Protostars and Planets VII}}, Vol. \bibinfo{volume}{534} of \emph{\bibinfo{series}{Astronomical Society of the Pacific Conference Series}}, \bibinfo{pages}{717} (\bibinfo{year}{2023}).
\newblock \eprint{2203.09759}.

\bibitem{Ansdell+2016}
\bibinfo{author}{{Ansdell}, M.} \emph{et~al.}
\newblock \bibinfo{title}{{ALMA Survey of Lupus Protoplanetary Disks. I. Dust and Gas Masses}}.
\newblock \emph{\bibinfo{journal}{\apj}} \textbf{\bibinfo{volume}{828}}, \bibinfo{pages}{46} (\bibinfo{year}{2016}).

\bibitem{Manara+18}
\bibinfo{author}{{Manara}, C.~F.}, \bibinfo{author}{{Morbidelli}, A.} \& \bibinfo{author}{{Guillot}, T.}
\newblock \bibinfo{title}{{Why do protoplanetary disks appear not massive enough to form the known exoplanet population?}}
\newblock \emph{\bibinfo{journal}{\aap}} \textbf{\bibinfo{volume}{618}}, \bibinfo{pages}{L3} (\bibinfo{year}{2018}).

\bibitem{Mulders+21}
\bibinfo{author}{{Mulders}, G.~D.}, \bibinfo{author}{{Pascucci}, I.}, \bibinfo{author}{{Ciesla}, F.~J.} \& \bibinfo{author}{{Fernandes}, R.~B.}
\newblock \bibinfo{title}{{The Mass Budgets and Spatial Scales of Exoplanet Systems and Protoplanetary Disks}}.
\newblock \emph{\bibinfo{journal}{\apj}} \textbf{\bibinfo{volume}{920}}, \bibinfo{pages}{66} (\bibinfo{year}{2021}).

\bibitem{Stefansson+23}
\bibinfo{author}{{Stefansson}, G.} \emph{et~al.}
\newblock \bibinfo{title}{{An extreme test case for planet formation: a close-in Neptune orbiting an ultracool star}}.
\newblock \emph{\bibinfo{journal}{arXiv e-prints}} \bibinfo{pages}{arXiv:2303.13321} (\bibinfo{year}{2023}).

\bibitem{Fedele+10}
\bibinfo{author}{{Fedele}, D.}, \bibinfo{author}{{van den Ancker}, M.~E.}, \bibinfo{author}{{Henning}, T.}, \bibinfo{author}{{Jayawardhana}, R.} \& \bibinfo{author}{{Oliveira}, J.~M.}
\newblock \bibinfo{title}{{Timescale of mass accretion in pre-main-sequence stars}}.
\newblock \emph{\bibinfo{journal}{\aap}} \textbf{\bibinfo{volume}{510}}, \bibinfo{pages}{A72} (\bibinfo{year}{2010}).

\bibitem{Manara+20}
\bibinfo{author}{{Manara}, C.~F.} \emph{et~al.}
\newblock \bibinfo{title}{{X-shooter survey of disk accretion in Upper Scorpius. I. Very high accretion rates at age > 5 Myr}}.
\newblock \emph{\bibinfo{journal}{\aap}} \textbf{\bibinfo{volume}{639}}, \bibinfo{pages}{A58} (\bibinfo{year}{2020}).

\bibitem{Mauco+23}
\bibinfo{author}{{Mauc{\'o}}, K.} \emph{et~al.}
\newblock \bibinfo{title}{{Testing external photoevaporation in the {\ensuremath{\sigma}}-Orionis cluster with spectroscopy and disk mass measurements}}.
\newblock \emph{\bibinfo{journal}{\aap}} \textbf{\bibinfo{volume}{679}}, \bibinfo{pages}{A82} (\bibinfo{year}{2023}).

\bibitem{Gaudel+20}
\bibinfo{author}{{Gaudel}, M.} \emph{et~al.}
\newblock \bibinfo{title}{{Angular momentum profiles of Class 0 protostellar envelopes}}.
\newblock \emph{\bibinfo{journal}{\aap}} \textbf{\bibinfo{volume}{637}}, \bibinfo{pages}{A92} (\bibinfo{year}{2020}).

\bibitem{Marino+15}
\bibinfo{author}{{Marino}, S.}, \bibinfo{author}{{Perez}, S.} \& \bibinfo{author}{{Casassus}, S.}
\newblock \bibinfo{title}{{Shadows Cast by a Warp in the HD 142527 Protoplanetary Disk}}.
\newblock \emph{\bibinfo{journal}{\apjl}} \textbf{\bibinfo{volume}{798}}, \bibinfo{pages}{L44} (\bibinfo{year}{2015}).

\bibitem{Pinilla+18}
\bibinfo{author}{{Pinilla}, P.} \emph{et~al.}
\newblock \bibinfo{title}{{Variable Outer Disk Shadowing around the Dipper Star RXJ1604.3-2130}}.
\newblock \emph{\bibinfo{journal}{\apj}} \textbf{\bibinfo{volume}{868}}, \bibinfo{pages}{85} (\bibinfo{year}{2018}).

\bibitem{Ansdell+20}
\bibinfo{author}{{Ansdell}, M.} \emph{et~al.}
\newblock \bibinfo{title}{{Are inner disc misalignments common? ALMA reveals an isotropic outer disc inclination distribution for young dipper stars}}.
\newblock \emph{\bibinfo{journal}{\mnras}} \textbf{\bibinfo{volume}{492}}, \bibinfo{pages}{572--588} (\bibinfo{year}{2020}).

\bibitem{Hubert+13}
\bibinfo{author}{{Huber}, D.} \emph{et~al.}
\newblock \bibinfo{title}{{Stellar Spin-Orbit Misalignment in a Multiplanet System}}.
\newblock \emph{\bibinfo{journal}{Science}} \textbf{\bibinfo{volume}{342}}, \bibinfo{pages}{331--334} (\bibinfo{year}{2013}).

\bibitem{Sanchis-Ojeda+13}
\bibinfo{author}{{Sanchis-Ojeda}, R.} \emph{et~al.}
\newblock \bibinfo{title}{{Kepler-63b: A Giant Planet in a Polar Orbit around a Young Sun-like Star}}.
\newblock \emph{\bibinfo{journal}{\apj}} \textbf{\bibinfo{volume}{775}}, \bibinfo{pages}{54} (\bibinfo{year}{2013}).

\bibitem{Kamiaka+19}
\bibinfo{author}{{Kamiaka}, S.} \emph{et~al.}
\newblock \bibinfo{title}{{The Misaligned Orbit of the Earth-sized Planet Kepler-408b}}.
\newblock \emph{\bibinfo{journal}{\aj}} \textbf{\bibinfo{volume}{157}}, \bibinfo{pages}{137} (\bibinfo{year}{2019}).

\bibitem{Hjorth+21}
\bibinfo{author}{{Hjorth}, M.} \emph{et~al.}
\newblock \bibinfo{title}{{A backward-spinning star with two coplanar planets}}.
\newblock \emph{\bibinfo{journal}{Proceedings of the National Academy of Science}} \textbf{\bibinfo{volume}{118}}, \bibinfo{pages}{e2017418118} (\bibinfo{year}{2021}).

\bibitem{Albrecht+22}
\bibinfo{author}{{Albrecht}, S.~H.}, \bibinfo{author}{{Dawson}, R.~I.} \& \bibinfo{author}{{Winn}, J.~N.}
\newblock \bibinfo{title}{{Stellar Obliquities in Exoplanetary Systems}}.
\newblock \emph{\bibinfo{journal}{\pasp}} \textbf{\bibinfo{volume}{134}}, \bibinfo{pages}{082001} (\bibinfo{year}{2022}).

\bibitem{Kawai+23}
\bibinfo{author}{{Kawai}, Y.}, \bibinfo{author}{{Narita}, N.}, \bibinfo{author}{{Fukui}, A.}, \bibinfo{author}{{Watanabe}, N.} \& \bibinfo{author}{{Inaba}, S.}
\newblock \bibinfo{title}{{The flipped orbit of KELT-19Ab inferred from the symmetric TESS transit light curves}}.
\newblock \emph{\bibinfo{journal}{\mnras}}  (\bibinfo{year}{2023}).

\bibitem{Pineda+20}
\bibinfo{author}{{Pineda}, J.~E.} \emph{et~al.}
\newblock \bibinfo{title}{{A protostellar system fed by a streamer of 10,500 au length}}.
\newblock \emph{\bibinfo{journal}{Nature Astronomy}}  (\bibinfo{year}{2020}).

\bibitem{Alves+20}
\bibinfo{author}{{Alves}, F.~O.} \emph{et~al.}
\newblock \bibinfo{title}{{A Case of Simultaneous Star and Planet Formation}}.
\newblock \emph{\bibinfo{journal}{\apjl}} \textbf{\bibinfo{volume}{904}}, \bibinfo{pages}{L6} (\bibinfo{year}{2020}).

\bibitem{Huang+21}
\bibinfo{author}{{Huang}, J.} \emph{et~al.}
\newblock \bibinfo{title}{{Molecules with ALMA at Planet-forming Scales (MAPS). XIX. Spiral Arms, a Tail, and Diffuse Structures Traced by CO around the GM Aur Disk}}.
\newblock \emph{\bibinfo{journal}{\apjs}} \textbf{\bibinfo{volume}{257}}, \bibinfo{pages}{19} (\bibinfo{year}{2021}).

\bibitem{Grant+21}
\bibinfo{author}{{Grant}, S.~L.} \emph{et~al.}
\newblock \bibinfo{title}{{An ALMA Survey of Protoplanetary Disks in Lynds 1641}}.
\newblock \emph{\bibinfo{journal}{\apj}} \textbf{\bibinfo{volume}{913}}, \bibinfo{pages}{123} (\bibinfo{year}{2021}).

\bibitem{Valdivia+22}
\bibinfo{author}{{Valdivia-Mena}, M.~T.} \emph{et~al.}
\newblock \bibinfo{title}{{PRODIGE - envelope to disk with NOEMA. I. A 3000 au streamer feeding a Class I protostar}}.
\newblock \emph{\bibinfo{journal}{\aap}} \textbf{\bibinfo{volume}{667}}, \bibinfo{pages}{A12} (\bibinfo{year}{2022}).

\bibitem{Cacciapuoti+23}
\bibinfo{author}{{Cacciapuoti}, L.} \emph{et~al.}
\newblock \bibinfo{title}{{A dusty streamer infalling onto the disk of a class I protostar. ALMA dual-band constraints on grain properties and mass infall rate}}.
\newblock \emph{\bibinfo{journal}{arXiv e-prints}} \bibinfo{pages}{arXiv:2311.13723} (\bibinfo{year}{2023}).

\bibitem{Hoyle+Lyttleton39}
\bibinfo{author}{{Hoyle}, F.} \& \bibinfo{author}{{Lyttleton}, R.~A.}
\newblock \bibinfo{title}{{The effect of interstellar matter on climatic variation}}.
\newblock \emph{\bibinfo{journal}{Proceedings of the Cambridge Philosophical Society}} \textbf{\bibinfo{volume}{35}}, \bibinfo{pages}{405} (\bibinfo{year}{1939}).

\bibitem{Bondi+Hoyle44}
\bibinfo{author}{{Bondi}, H.} \& \bibinfo{author}{{Hoyle}, F.}
\newblock \bibinfo{title}{{On the mechanism of accretion by stars}}.
\newblock \emph{\bibinfo{journal}{\mnras}} \textbf{\bibinfo{volume}{104}}, \bibinfo{pages}{273} (\bibinfo{year}{1944}).

\bibitem{Bondi52}
\bibinfo{author}{{Bondi}, H.}
\newblock \bibinfo{title}{{On spherically symmetrical accretion}}.
\newblock \emph{\bibinfo{journal}{\mnras}} \textbf{\bibinfo{volume}{112}}, \bibinfo{pages}{195} (\bibinfo{year}{1952}).

\bibitem{Padoan+05_BH}
\bibinfo{author}{{Padoan}, P.}, \bibinfo{author}{{Kritsuk}, A.}, \bibinfo{author}{{Norman}, M.~L.} \& \bibinfo{author}{{Nordlund}, {\AA}.}
\newblock \bibinfo{title}{{A Solution to the Pre-Main-Sequence Accretion Problem}}.
\newblock \emph{\bibinfo{journal}{\apjl}} \textbf{\bibinfo{volume}{622}}, \bibinfo{pages}{L61--L64} (\bibinfo{year}{2005}).

\bibitem{Throop+Bally08}
\bibinfo{author}{{Throop}, H.~B.} \& \bibinfo{author}{{Bally}, J.}
\newblock \bibinfo{title}{{Tail-End Bondi-Hoyle Accretion in Young Star Clusters: Implications for Disks, Planets, and Stars}}.
\newblock \emph{\bibinfo{journal}{\aj}} \textbf{\bibinfo{volume}{135}}, \bibinfo{pages}{2380--2397} (\bibinfo{year}{2008}).

\bibitem{Padoan+14acc}
\bibinfo{author}{{Padoan}, P.}, \bibinfo{author}{{Haugb{\o}lle}, T.} \& \bibinfo{author}{{Nordlund}, {\AA}.}
\newblock \bibinfo{title}{{Infall-driven Protostellar Accretion and the Solution to the Luminosity Problem}}.
\newblock \emph{\bibinfo{journal}{\apj}} \textbf{\bibinfo{volume}{797}}, \bibinfo{pages}{32} (\bibinfo{year}{2014}).

\bibitem{Andrews+2018}
\bibinfo{author}{{Andrews}, S.~M.} \emph{et~al.}
\newblock \bibinfo{title}{{The Disk Substructures at High Angular Resolution Project (DSHARP). I. Motivation, Sample, Calibration, and Overview}}.
\newblock \emph{\bibinfo{journal}{\apjl}} \textbf{\bibinfo{volume}{869}}, \bibinfo{pages}{L41} (\bibinfo{year}{2018}).

\bibitem{Hendler+20}
\bibinfo{author}{{Hendler}, N.} \emph{et~al.}
\newblock \bibinfo{title}{{The Evolution of Dust Disk Sizes from a Homogeneous Analysis of 1-10 Myr old Stars}}.
\newblock \emph{\bibinfo{journal}{\apj}} \textbf{\bibinfo{volume}{895}}, \bibinfo{pages}{126} (\bibinfo{year}{2020}).

\bibitem{Andrews20}
\bibinfo{author}{{Andrews}, S.~M.}
\newblock \bibinfo{title}{{Observations of Protoplanetary Disk Structures}}.
\newblock \emph{\bibinfo{journal}{\araa}} \textbf{\bibinfo{volume}{58}}, \bibinfo{pages}{483--528} (\bibinfo{year}{2020}).

\bibitem{Long+22}
\bibinfo{author}{{Long}, F.} \emph{et~al.}
\newblock \bibinfo{title}{{Gas Disk Sizes from CO Line Observations: A Test of Angular Momentum Evolution}}.
\newblock \emph{\bibinfo{journal}{\apj}} \textbf{\bibinfo{volume}{931}}, \bibinfo{pages}{6} (\bibinfo{year}{2022}).

\bibitem{Testi+22}
\bibinfo{author}{{Testi}, L.} \emph{et~al.}
\newblock \bibinfo{title}{{The protoplanetary disk population in the {\ensuremath{\rho}}-Ophiuchi region L1688 and the time evolution of Class II YSOs}}.
\newblock \emph{\bibinfo{journal}{\aap}} \textbf{\bibinfo{volume}{663}}, \bibinfo{pages}{A98} (\bibinfo{year}{2022}).

\bibitem{Kuffmeier+23}
\bibinfo{author}{{Kuffmeier}, M.}, \bibinfo{author}{{Jensen}, S.~S.} \& \bibinfo{author}{{Haugb{\o}lle}, T.}
\newblock \bibinfo{title}{{Rejuvenating infall: a crucial yet overlooked source of mass and angular momentum}}.
\newblock \emph{\bibinfo{journal}{European Physical Journal Plus}} \textbf{\bibinfo{volume}{138}}, \bibinfo{pages}{272} (\bibinfo{year}{2023}).

\bibitem{Goodman+93}
\bibinfo{author}{{Goodman}, A.~A.}, \bibinfo{author}{{Benson}, P.~J.}, \bibinfo{author}{{Fuller}, G.~A.} \& \bibinfo{author}{{Myers}, P.~C.}
\newblock \bibinfo{title}{Dense cores in dark clouds. viii - velocity gradients}.
\newblock \emph{\bibinfo{journal}{ApJ}} \textbf{\bibinfo{volume}{406}}, \bibinfo{pages}{528} (\bibinfo{year}{1993}).

\bibitem{Phillips99}
\bibinfo{author}{{Phillips}, J.~P.}
\newblock \bibinfo{title}{{Rotation in molecular clouds}}.
\newblock \emph{\bibinfo{journal}{\aaps}} \textbf{\bibinfo{volume}{134}}, \bibinfo{pages}{241--254} (\bibinfo{year}{1999}).

\bibitem{Imara+Blitz2011}
\bibinfo{author}{{Imara}, N.} \& \bibinfo{author}{{Blitz}, L.}
\newblock \bibinfo{title}{{Angular Momentum in Giant Molecular Clouds. I. The Milky Way}}.
\newblock \emph{\bibinfo{journal}{\apj}} \textbf{\bibinfo{volume}{732}}, \bibinfo{pages}{78} (\bibinfo{year}{2011}).

\bibitem{Tatematsu+16}
\bibinfo{author}{{Tatematsu}, K.} \emph{et~al.}
\newblock \bibinfo{title}{{Angular momentum of the N$_{2}$H$^{+}$ cores in the Orion A cloud}}.
\newblock \emph{\bibinfo{journal}{\pasj}} \textbf{\bibinfo{volume}{68}}, \bibinfo{pages}{24} (\bibinfo{year}{2016}).

\bibitem{Cieza+21}
\bibinfo{author}{{Cieza}, L.~A.} \emph{et~al.}
\newblock \bibinfo{title}{{The Ophiuchus DIsc Survey Employing ALMA (ODISEA) - III. The evolution of substructures in massive discs at 3-5 au resolution}}.
\newblock \emph{\bibinfo{journal}{\mnras}} \textbf{\bibinfo{volume}{501}}, \bibinfo{pages}{2934--2953} (\bibinfo{year}{2021}).

\bibitem{Stapper+22}
\bibinfo{author}{{Stapper}, L.~M.}, \bibinfo{author}{{Hogerheijde}, M.~R.}, \bibinfo{author}{{van Dishoeck}, E.~F.} \& \bibinfo{author}{{Mentel}, R.}
\newblock \bibinfo{title}{{The mass and size of Herbig disks as seen by ALMA}}.
\newblock \emph{\bibinfo{journal}{\aap}} \textbf{\bibinfo{volume}{658}}, \bibinfo{pages}{A112} (\bibinfo{year}{2022}).

\bibitem{Pelkonen+24}
\bibinfo{author}{{Pelkonen}, V.~M.}, \bibinfo{author}{{Padoan}, P.}, \bibinfo{author}{{Juvela}, M.}, \bibinfo{author}{{Haugb{\o}lle}, T.} \& \bibinfo{author}{{Nordlund}, {\r{A}}.}
\newblock \bibinfo{title}{{Origin and Evolution of Angular Momentum of Class II Disks}}.
\newblock \emph{\bibinfo{journal}{A\&A}}
\textbf{\bibinfo{volume}{694}}, \bibinfo{pages}{A327}(\bibinfo{year}{2025}).

\bibitem{Manara+23}
\bibinfo{author}{{Manara}, C.~F.} \emph{et~al.}
\newblock \bibinfo{editor}{{Inutsuka}, S.}, \bibinfo{editor}{{Aikawa}, Y.}, \bibinfo{editor}{{Muto}, T.}, \bibinfo{editor}{{Tomida}, K.} \& \bibinfo{editor}{{Tamura}, M.} (eds) \emph{\bibinfo{title}{{Demographics of Young Stars and their Protoplanetary Disks: Lessons Learned on Disk Evolution and its Connection to Planet Formation}}}.
\newblock (eds \bibinfo{editor}{{Inutsuka}, S.}, \bibinfo{editor}{{Aikawa}, Y.}, \bibinfo{editor}{{Muto}, T.}, \bibinfo{editor}{{Tomida}, K.} \& \bibinfo{editor}{{Tamura}, M.}) \emph{\bibinfo{booktitle}{Protostars and Planets VII}}, Vol. \bibinfo{volume}{534} of \emph{\bibinfo{series}{Astronomical Society of the Pacific Conference Series}}, \bibinfo{pages}{539} (\bibinfo{year}{2023}).
\newblock \eprint{2203.09930}.

\bibitem{Beltran+deWit16}
\bibinfo{author}{{Beltr{\'a}n}, M.~T.} \& \bibinfo{author}{{de Wit}, W.~J.}
\newblock \bibinfo{title}{{Accretion disks in luminous young stellar objects}}.
\newblock \emph{\bibinfo{journal}{\aapr}} \textbf{\bibinfo{volume}{24}}, \bibinfo{pages}{6} (\bibinfo{year}{2016}).

\bibitem{Gupta+23}
\bibinfo{author}{{Gupta}, A.} \emph{et~al.}
\newblock \bibinfo{title}{{Reflections on nebulae around young stars. A systematic search for late-stage infall of material onto Class II disks}}.
\newblock \emph{\bibinfo{journal}{\aap}} \textbf{\bibinfo{volume}{670}}, \bibinfo{pages}{L8} (\bibinfo{year}{2023}).

\bibitem{Kuffmeier+17}
\bibinfo{author}{{Kuffmeier}, M.}, \bibinfo{author}{{Haugb{\o}lle}, T.} \& \bibinfo{author}{{Nordlund}, {\r{A}}.}
\newblock \bibinfo{title}{{Zoom-in Simulations of Protoplanetary Disks Starting from GMC Scales}}.
\newblock \emph{\bibinfo{journal}{\apj}} \textbf{\bibinfo{volume}{846}}, \bibinfo{pages}{7} (\bibinfo{year}{2017}).

\bibitem{Kuffmeier+20}
\bibinfo{author}{{Kuffmeier}, M.}, \bibinfo{author}{{Goicovic}, F.~G.} \& \bibinfo{author}{{Dullemond}, C.~P.}
\newblock \bibinfo{title}{{Late encounter events as source of disks and spiral structures. Forming second generation disks}}.
\newblock \emph{\bibinfo{journal}{\aap}} \textbf{\bibinfo{volume}{633}}, \bibinfo{pages}{A3} (\bibinfo{year}{2020}).

\bibitem{Kuffmeier+21}
\bibinfo{author}{{Kuffmeier}, M.}, \bibinfo{author}{{Dullemond}, C.~P.}, \bibinfo{author}{{Reissl}, S.} \& \bibinfo{author}{{Goicovic}, F.~G.}
\newblock \bibinfo{title}{{Misaligned disks induced by infall}}.
\newblock \emph{\bibinfo{journal}{\aap}} \textbf{\bibinfo{volume}{656}}, \bibinfo{pages}{A161} (\bibinfo{year}{2021}).

\bibitem{Rabatin+Collins23}
\bibinfo{author}{{Rabatin}, B.} \& \bibinfo{author}{{Collins}, D.~C.}
\newblock \bibinfo{title}{{Density and velocity correlations in isothermal supersonic turbulence}}.
\newblock \emph{\bibinfo{journal}{\mnras}} \textbf{\bibinfo{volume}{525}}, \bibinfo{pages}{297--310} (\bibinfo{year}{2023}).

\bibitem{Larson81}
\bibinfo{author}{{Larson}, R.~B.}
\newblock \bibinfo{title}{{Turbulence and star formation in molecular clouds}}.
\newblock \emph{\bibinfo{journal}{MNRAS}} \textbf{\bibinfo{volume}{194}}, \bibinfo{pages}{809--826} (\bibinfo{year}{1981}).
\newblock \urlprefix\url{http://adsabs.harvard.edu/cgi-bin/nph-bib_query?bibcode=1981MNRAS.194..809L&db_key=AST}.

\bibitem{Solomon+87}
\bibinfo{author}{Solomon, P.~M.}, \bibinfo{author}{Rivolo, A.~R.}, \bibinfo{author}{Barrett, J.~W.} \& \bibinfo{author}{Yahil, A.~M.}
\newblock \emph{\bibinfo{journal}{ApJ}} \textbf{\bibinfo{volume}{319}}, \bibinfo{pages}{730} (\bibinfo{year}{1987}).

\bibitem{Jorgensen+22}
\bibinfo{author}{{J{\o}rgensen}, J.~K.} \emph{et~al.}
\newblock \bibinfo{title}{{Binarity of a protostar affects the evolution of the disk and planets}}.
\newblock \emph{\bibinfo{journal}{\nat}} \textbf{\bibinfo{volume}{606}}, \bibinfo{pages}{272--275} (\bibinfo{year}{2022}).

\bibitem{Jensen+23}
\bibinfo{author}{{Jensen}, S.~S.}, \bibinfo{author}{{Spezzano}, S.}, \bibinfo{author}{{Caselli}, P.}, \bibinfo{author}{{Grassi}, T.} \& \bibinfo{author}{{Haugb{\o}lle}, T.}
\newblock \bibinfo{title}{{3D physico-chemical model of a pre-stellar core. I. Environmental and structural impact on the distribution of CH$_{3}$OH and c-C$_{3}$H$_{2}$}}.
\newblock \emph{\bibinfo{journal}{\aap}} \textbf{\bibinfo{volume}{675}}, \bibinfo{pages}{A34} (\bibinfo{year}{2023}).

\bibitem{Kuffmeier+24}
\bibinfo{author}{{Kuffmeier}, M.}, \bibinfo{author}{{Pineda}, J.~E.}, \bibinfo{author}{{Segura-Cox}, D.} \& \bibinfo{author}{{Haugb{\o}lle}, T.}
\newblock \bibinfo{title}{{Constraints on the (re-)orientation of star-disk systems through infall}}.
\newblock \emph{\bibinfo{journal}{A\&A (submitted)}}  (\bibinfo{year}{2024}).

\bibitem{Haugbolle+18imf}
\bibinfo{author}{{Haugb{\o}lle}, T.}, \bibinfo{author}{{Padoan}, P.} \& \bibinfo{author}{{Nordlund}, {\r{A}}.}
\newblock \bibinfo{title}{{The Stellar IMF from Isothermal MHD Turbulence}}.
\newblock \emph{\bibinfo{journal}{\apj}} \textbf{\bibinfo{volume}{854}}, \bibinfo{pages}{35} (\bibinfo{year}{2018}).

\bibitem{Teyssier07}
\bibinfo{author}{{Teyssier}, R.}
\newblock \bibinfo{title}{{A high order Godunov scheme with constrained transport and adaptive mesh refinement for astrophysical and geophysical MHD}}.
\newblock \emph{\bibinfo{journal}{Geophysical and Astrophysical Fluid Dynamics}} \textbf{\bibinfo{volume}{101}}, \bibinfo{pages}{199--225} (\bibinfo{year}{2007}).

\bibitem{Salpeter55}
\bibinfo{author}{{Salpeter}, E.~E.}
\newblock \bibinfo{title}{The luminosity function and stellar evolution.}
\newblock \emph{\bibinfo{journal}{ApJ}} \textbf{\bibinfo{volume}{121}}, \bibinfo{pages}{161} (\bibinfo{year}{1955}).
\newblock \urlprefix\url{http://adsabs.harvard.edu/cgi-bin/nph-bib_query?bibcode=1955ApJ...121..161S&db_key=AST}.

\bibitem{Chabrier05}
\bibinfo{author}{{Chabrier}, G.}
\newblock \bibinfo{editor}{{E.~Corbelli, F.~Palla, \& H.~Zinnecker}} (ed.) \emph{\bibinfo{title}{{The Initial Mass Function: from Salpeter 1955 to 2005}}}.
\newblock (ed.\bibinfo{editor}{{E.~Corbelli, F.~Palla, \& H.~Zinnecker}}) \emph{\bibinfo{booktitle}{The Initial Mass Function 50 Years Later}}, Vol. \bibinfo{volume}{327} of \emph{\bibinfo{series}{Astrophysics and Space Science Library}}, \bibinfo{pages}{41--+} (\bibinfo{year}{2005}).
\newblock \eprint{arXiv:astro-ph/0409465}.

\bibitem{Weidenschilling77}
\bibinfo{author}{{Weidenschilling}, S.~J.}
\newblock \bibinfo{title}{{The Distribution of Mass in the Planetary System and Solar Nebula}}.
\newblock \emph{\bibinfo{journal}{\apss}} \textbf{\bibinfo{volume}{51}}, \bibinfo{pages}{153--158} (\bibinfo{year}{1977}).

\bibitem{Padoan+20massive}
\bibinfo{author}{{Padoan}, P.}, \bibinfo{author}{{Pan}, L.}, \bibinfo{author}{{Juvela}, M.}, \bibinfo{author}{{Haugb{\o}lle}, T.} \& \bibinfo{author}{{Nordlund}, {\r{A}}.}
\newblock \bibinfo{title}{{The Origin of Massive Stars: The Inertial-inflow Model}}.
\newblock \emph{\bibinfo{journal}{\apj}} \textbf{\bibinfo{volume}{900}}, \bibinfo{pages}{82} (\bibinfo{year}{2020}).

\bibitem{Pelkonen+21}
\bibinfo{author}{{Pelkonen}, V.~M.}, \bibinfo{author}{{Padoan}, P.}, \bibinfo{author}{{Haugb{\o}lle}, T.} \& \bibinfo{author}{{Nordlund}, {\r{A}}.}
\newblock \bibinfo{title}{{From the CMF to the IMF: beyond the core-collapse model}}.
\newblock \emph{\bibinfo{journal}{\mnras}} \textbf{\bibinfo{volume}{504}}, \bibinfo{pages}{1219--1236} (\bibinfo{year}{2021}).

\bibitem{Krumholz+05BH}
\bibinfo{author}{{Krumholz}, M.~R.}, \bibinfo{author}{{McKee}, C.~F.} \& \bibinfo{author}{{Klein}, R.~I.}
\newblock \bibinfo{title}{{Bondi Accretion in the Presence of Vorticity}}.
\newblock \emph{\bibinfo{journal}{\apj}} \textbf{\bibinfo{volume}{618}}, \bibinfo{pages}{757--768} (\bibinfo{year}{2005}).

\bibitem{Krumholz+06BH}
\bibinfo{author}{{Krumholz}, M.~R.}, \bibinfo{author}{{McKee}, C.~F.} \& \bibinfo{author}{{Klein}, R.~I.}
\newblock \bibinfo{title}{{Bondi-Hoyle Accretion in a Turbulent Medium}}.
\newblock \emph{\bibinfo{journal}{\apj}} \textbf{\bibinfo{volume}{638}}, \bibinfo{pages}{369--381} (\bibinfo{year}{2006}).

\bibitem{Lannelongue+21_green}
\bibinfo{author}{Lannelongue, L.}, \bibinfo{author}{Grealey, J.} \& \bibinfo{author}{Inouye, M.}
\newblock \bibinfo{title}{Green algorithms: Quantifying the carbon footprint of computation}.
\newblock \emph{\bibinfo{journal}{Advanced Science}} \textbf{\bibinfo{volume}{8}}, \bibinfo{pages}{2100707} (\bibinfo{year}{2021}).
\newblock \urlprefix\url{https://onlinelibrary.wiley.com/doi/abs/10.1002/advs.202100707}.

\end{thebibliography}

\end{document}